\documentclass[superscriptaddress,
 reprint,
 amssymb, amsmath,
 aip, cha
]{revtex4-1}
\usepackage[utf8]{inputenc}
\usepackage[T1]{fontenc}
\usepackage{amsmath}
\usepackage{amsfonts}
\usepackage{amssymb}
\usepackage{graphicx}
\usepackage{bm}
\usepackage{grffile} 
\usepackage{xr} 
\usepackage{xcolor}
\usepackage{mathtools}
\usepackage{amsthm}
\usepackage{enumerate}

\graphicspath{{figures/}}

\newcommand{\KgKM}{K_g^{\text{KM}}}
\newcommand{\KgCC}{K_g^{\text{CC}}}

\newtheorem{defn}{Definition}

\begin{document}


\title{Mesoscopic model reduction for the collective dynamics of sparse coupled oscillator networks}

\author{Lauren~D. Smith}
 \email{lauren.smith@auckland.ac.nz}
  \affiliation{\mbox{Department of Mathematics, The University of Auckland, Auckland, 1142, New Zealand}}
   \affiliation{\mbox{School of Mathematics and Statistics, The University of Sydney, Sydney, NSW 2006, Australia}}
\author{Georg~A. Gottwald}
 \email{georg.gottwald@sydney.edu.au}
 \affiliation{\mbox{School of Mathematics and Statistics, The University of Sydney, Sydney, NSW 2006, Australia}}

\date{\today}

\begin{abstract}
The behavior at bifurcation from global synchronization to partial synchronization in finite networks of coupled oscillators is a complex phenomenon, involving the intricate dynamics of one or more oscillators with the remaining synchronized oscillators. This is not captured well by standard macroscopic model reduction techniques which capture only the collective behavior of synchronized oscillators in the thermodynamic limit. We introduce two mesoscopic model reductions for finite sparse networks of coupled oscillators to quantitatively capture the dynamics close to bifurcation from global to partial synchronization. Our model reduction builds upon the method of collective coordinates. We first show that standard collective coordinate reduction has difficulties capturing this bifurcation. We identify a particular topological structure at bifurcation consisting of a main synchronized cluster, the oscillator that desynchronizes at bifurcation, and an intermediary node connecting them. Utilizing this structure and ensemble averages we derive an analytic expression for the mismatch between the true bifurcation from global to partial synchronization and its estimate calculated via the collective coordinate approach. This allows to calibrate the standard collective coordinate approach without prior knowledge of which node will desynchronize. We introduce a second mesoscopic reduction, utilizing the same particular topological structure, which allows for a quantitative dynamical description of the phases near bifurcation. The mesoscopic reductions significantly reduce the computational complexity of the collective coordinate approach, reducing from $\mathcal{O}(N^2)$ to $\mathcal{O}(1)$. We perform numerical simulations for Erd\H os-R\'enyi networks and for modified Barab\'asi-Albert networks demonstrating excellent quantitative agreement at and close to bifurcation.
\end{abstract}

\maketitle

\begin{quotation}
Model reduction is essential to gaining a deeper understanding of the dynamics of complex systems. For coupled oscillator networks, full models have very high dimension, but the resultant dynamics is often low-dimensional. For example, synchronization is a collective phenomenon that can be described by a small number of variables. There exist well-established model reduction methods for networks with infinitely many oscillators, but few methods for finite networks. Here we consider the collective coordinate model reduction approach applied to sparsely connected finite networks. We show that while the standard collective approach yields significant error for these sparse networks, this error can be corrected by considering mesoscopic reductions that capture the microscopic detail of essential nodes but only the macroscopic detail of the majority of the network. Our reductions yield a correction to the critical coupling strength $K_g$, corresponding to the transition from global to partial synchronization, obtained via the standard collective coordinate approach, as well as highly simplified but highly accurate temporal dynamics of individual oscillators.
\end{quotation}

\section{Introduction}

Synchronization is common to many networks of coupled oscillators, including in natural systems such as the activity of the brain \cite{SheebaEtAl08, BhowmikShanahan12} and synchronous firefly flashing \cite{MirolloStrogatz90}, as well as in many engineering applications, such as power grids \cite{FilatrellaEtAl08, NishikawaMotter2015},  and Josephson junction arrays  \cite{WatanabeStrogatz94, WiesenfeldEtAl98}. Understanding the transition from global synchronization, such that all oscillators are synchronized, to partial synchronization, with only some oscillators synchronized, or complete incoherence is essential to control the collective behavior of complex networks of oscillators. For example, in the case of power grids, all oscillators need to remain synchronized, otherwise blackouts occur, and identifying which parts of the power grid are most prone to desynchronization is of particular importance in controlling them. 

The high dimensionality of coupled oscillator networks makes detailed analysis intractable. As such, several model reduction methods have been developed that reduce the dimension of coupled oscillator systems. Here we consider the Kuramoto model \cite{Kuramoto84, Strogatz00, PikovskyEtAl01, AcebronEtAl05, OsipovEtAl07, ArenasEtAl08, DorflerBullo14, RodriguesEtAl16} for coupled oscillators. In the thermodynamic limit of infinitely many oscillators, dimension reduction can be achieved using the Ott-Antonsen ansatz \cite{OttAntonsen08}, which describes the dynamics of the macroscopic order parameter. For finite networks, the collective coordinate method \cite{Gottwald15, Gottwald17, HancockGottwald18, SmithGottwald19, SmithGottwald20, YueEtAl20} achieves dimension reduction by projecting the dynamics of the full system onto a judiciously chosen ansatz manifold, yielding evolution equations for macroscopic variables such as the order parameter. It has recently been shown that in the thermodynamic limit there is an equivalence between the collective coordinate approach and the Ott-Antonsen approach \cite{SmithGottwald20}.

The collective coordinate approach is designed to describe the collective dynamics of a synchronized cluster of oscillators. Close to the bifurcation from global synchronization to partial synchronization, the dynamics of individual oscillators or of a group of oscillators which are about to break off from the main synchronized cluster becomes important. The dynamics of these most unstable oscillators is not captured by the standard collective coordinate framework, and, hence the standard collective coordinate approach provides a less accurate quantitative approximation close to the bifurcation. The loss of accuracy of the standard collective coordinate approach in describing the bifurcation point $K_g$ is much more pronounced in sparse networks compared to dense networks. We will show that the estimate of $K_g$ itself may incur an error of $10-25\%$ for highly connected networks and $25-35\%$ for sparsely connected networks. We therefore focus here on sparsely connected networks, such that the mean degree $k$ is significantly smaller than the size of the network $N$.

\begin{figure}[tbp]
\centering
\includegraphics[width=0.8\columnwidth]{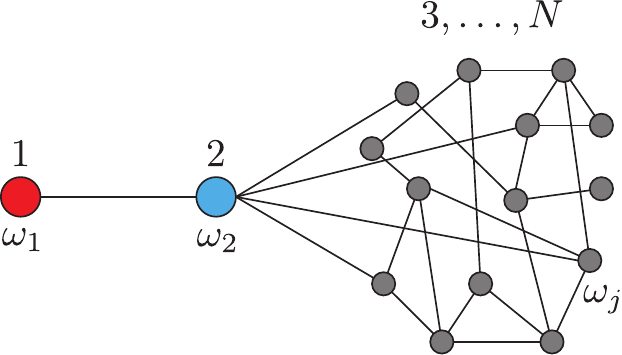}
\caption{
The network structure assumed here for mesoscopic reduction of sparse networks: a degree one node (labeled node 1, red), connects to node 2 (blue), which connects randomly to the rest of the network (gray).
}
\label{fig:degree_one_network_schematic}
\end{figure}

To improve the quantitative accuracy of the collective coordinate method near the bifurcation at $K=K_g$ we propose two different mesoscopic collective coordinate reductions. Both mesoscopic approaches assume that the network has the structure shown in Fig.~\ref{fig:degree_one_network_schematic}, such that node 1 is the most susceptible to desynchronization and has degree one. Node 1 connects to node 2, which connects randomly to the rest of the network. This structure is common for sparse networks where there are many degree one nodes, and the most susceptible node is generally the one with natural frequency furthest from the mean frequency, i.e., $|\omega_i - \Omega| \gg 1$. Our reductions are mesoscopic in the sense that they incorporate the microscopic detail of nodes 1 and 2, but only the macroscopic detail of the rest of the network which remains strongly synchronized. We remark that degree 1 nodes have recently been identified as being crucial in the stability of realistic power grids, where they are coined dead ends \cite{MenckEtAl14}.

The first mesoscopic collective coordinate approach involves a single collective coordinate and a one-dimensional (1D) ansatz manifold. The mesoscopic description yields an analytic expression that accurately captures the approximately linear relationship we observe between the actual values of the critical coupling strength $K_g$ and the estimates obtained via the standard collective coordinate approach. The analytic expression obtained depends only on ensemble statistical parameters of the full system, such as the mean degree $k$ and the variance of the natural frequencies $\sigma^2$. Therefore, even without any prior knowledge of which node is first to desynchronize, we obtain a highly accurate estimate for $K_g$ through correcting the standard collective coordinate approach.

The second mesoscopic collective coordinate approach uses four collective coordinates to capture the microscopic dynamics of two critical nodes (nodes 1 and 2 in Fig.~\ref{fig:degree_one_network_schematic}) together with the macroscopic dynamics of rest of the network. This yields a significantly simplified system that accurately captures the dynamics of the full system for coupling strengths $K$ close too the critical coupling strength $K_g$. Identifying the two critical nodes requires some prior knowledge of the microscopic dynamics of the full system, specifically, the node that is most susceptible to desynchronization which will be the first to break off from the synchronized cluster. To identify this node, one can use the standard collective framework \cite{HancockGottwald18}. The resulting simplified four-dimensional mesoscopic system accurately describes the temporal phase dynamics of the full system for coupling strengths $K>K_g$, such that there is convergence to a globally synchronized stated, and for $K<K_g$, such that there is non-stationary dynamics.

The paper is organized as follows. In Section~\ref{sec:KM_model} we discuss the Kuramoto model, its bifurcation for networks of the form Fig.~\ref{fig:degree_one_network_schematic}, and the two types of network structure we consider, Erd\H os-R\'enyi networks and modified Barab\'asi-Albert networks. In Section~\ref{sec:standard_CC} we describe standard collective coordinate model reduction. In Section~\ref{sec:single_cluster_simplified} we describe a mesoscopic collective coordinate reduction with one collective coordinate, which results in an analytic expression for the relationship between the actual and estimated values of the critical coupling strength $K_g$, and, hence, a means to correct the standard collective coordinate approach. In Section~\ref{sec:three_cluster} we describe a mesoscopic collective coordinate approach with four collective coordinates that we show accurately describes the temporal phase dynamics of the full system. Lastly, in Section~\ref{sec:conclusions} we summarize our results.

\section{The model} \label{sec:KM_model}

In the widely-studied Kuramoto model \cite{Kuramoto84, Strogatz00, PikovskyEtAl01, AcebronEtAl05, OsipovEtAl07, ArenasEtAl08, DorflerBullo14, RodriguesEtAl16} the dynamics of each oscillator $i$ with phase $\phi_i$ is governed by
\begin{equation} \label{eq:full_KM} 
\dot{\phi_i} = \omega_i + \frac{K}{N} \sum_{j=1}^N A_{ij} \sin(\phi_j - \phi_i),
\end{equation}
where $\omega_i$ is the natural frequency, drawn from a probability distribution $g(\omega)$, $K$ is the coupling strength, $N$ is the total number of oscillators and $A$ is the network adjacency matrix. Note that through a change of coordinates into a rotating reference frame, $\phi_i(t) \to \phi_i(t) - \Omega \, t$, where $\Omega = \frac{1}{N} \sum \omega_i$ is the mean frequency, we may assume without loss of generality that the mean frequency $\Omega$ is zero. Here we consider normally distributed natural frequencies $g(\omega) \sim \mathcal{N}(0,\sigma^2)$, and in all computations we take $\sigma^2 = 0.1$.

We now analyze the degree one break-off of node 1 for networks of the form Fig.~\ref{fig:degree_one_network_schematic}. The evolution equation for $\phi_1$ becomes
\begin{equation}
\dot\phi_1 = \omega_1 + \frac{K}{N} \sin(\phi_2 - \phi_1),
\end{equation}
and so it is clear that two solutions to $\dot \phi_1 = 0$ exist for $K > K_g = |\omega_1| N$, namely 
\begin{equation}
\phi_2 - \phi_1 = \arcsin\left(\frac{\omega_1 N}{K}\right),\,  \pi - \arcsin\left(\frac{\omega_1 N}{K}\right).
\end{equation}
At $K = K_g$ these two solutions coincide, both equaling $\pi/2$, and for $K<K_g$ there are no solutions to $\dot\phi_1 = 0$, indicating that $\phi_1$ is no longer synchronized. This transition describes the saddle-node bifurcation from global synchronization to partial synchronization that occurs at $K=K_g = |\omega_1| N$. This can also be shown by considering the Kuramoto model (\ref{eq:full_KM}) in phase difference coordinates $\Phi_i = \phi_{i+1} - \phi_i$ for $i = 1,\dots, N-1$. Under this change of coordinates and with the network structure in Fig.~\ref{fig:degree_one_network_schematic} the Kuramoto model becomes
\begin{align}
 \dot{\Phi}_1 &= \Delta \omega_1 + \frac{K}{N} \left( -2\sin\Phi_1  + F_1\left(\Phi_2, \dots, \Phi_{N-1} \right) \right)   \nonumber \\
 \dot{\Phi}_2 &= \Delta \omega_2 + \frac{K}{N} \left( \sin\Phi_1  + F_2\left(\Phi_2, \dots, \Phi_{N-1} \right) \right)   \label{eq:differenced_KM} \\
 \dot{\Phi}_j &= \Delta \omega_j + \frac{K}{N} F_j\left(\Phi_2, \dots, \Phi_{N-1} \right), \quad j=3,\dots,N-1,  \nonumber
\end{align}
where $\Delta \omega_i = \omega_{i+1} - \omega_i$, and each function $F_i$ is a linear combination of sine functions of the form $\sin(\sum_{j=a}^b  \Phi_j )$. The important feature is that the functions $F_i$ do not depend on $\Phi_1$. Linearizing around $\bm{\Phi} = 0$, the Jacobian of this differenced system is
\begin{equation} \label{eq:differenced_Jacobian}
J = \frac{K}{N} \left(
\begin{matrix}
-2 \cos \Phi_1 & \frac{\partial F_1}{\partial \Phi_2} & \frac{\partial F_1}{\partial \Phi_3} & \hdots \\
\cos \Phi_1 & \frac{\partial F_2}{\partial \Phi_2} & \frac{\partial F_2}{\partial \Phi_3} & \hdots \\
0 & \frac{\partial F_3}{\partial \Phi_2} & \frac{\partial F_3}{\partial \Phi_3} & \hdots \\
\vdots & \vdots & \vdots & \ddots
\end{matrix}
\right).
\end{equation}
At $\Phi_1 = \pi/2$, $J$ has a zero eigenvalue with corresponding eigenvector $(1,0,0,\dots)$, and $\det J$ changes sign crossing the hyperplane $\Phi_1 = \pi/2$, meaning at least one eigenvalue changes sign and a saddle-node bifurcation has occurred. This description using the differenced system (\ref{eq:differenced_KM}) and its Jacobian (\ref{eq:differenced_Jacobian}) will be essential to the mesoscopic collective coordinate approach described in Section~\ref{sec:single_cluster_simplified}.\\

We now describe the two types of networks used in our computations and analysis, Erd\H os-R\'enyi networks and modified Barab\'asi-Albert networks.

\subsection{Erd\H os-R\'enyi networks}

In an Erd\H os-R\'enyi (ER) network \cite{ErdosRenyi60}, each node in a simply connected network is connected to each other node with probability $0<p\leq 1$. Therefore, the mean degree of the network is $k = (N-1) p$. For $p\approx 1$ the network is dense, with $p=1$ corresponding to all-to-all coupling. For sufficiently small $p$, the network is sparse and there are likely to be several degree one nodes leading to a structure as in Fig.~\ref{fig:degree_one_network_schematic}.

For sparse networks with $p \ll 1$ the minimum degree of the network can be greater than one. We will show that our method for correcting $K_g$ obtained via the standard collective coordinate approach still applies to these networks with minimum degree greater than one, and which do not have the form shown in Fig.~\ref{fig:degree_one_network_schematic}.

\subsection{Modified Barab\'asi-Albert networks}

\begin{figure}[tbp]
\centering
\includegraphics[width=0.9\columnwidth]{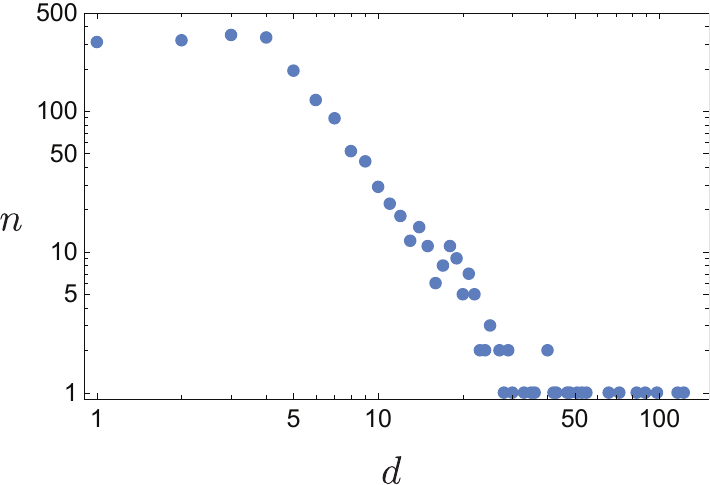}
\caption{
Degree distribution for a modified Barab\'asi-Albert network with $m_1 = 1$ and $m_2 = 4$ for $N=2000$ nodes. The graph uses log-log axes to highlight the power law scaling $n \sim d^a$.
}
\label{fig:modified_BA_degree_distribution}
\end{figure}

Barab\'asi-Albert (BA) networks \cite{BarabasiAlbert99} are scale-free graphs generated through preferential attachment. These graphs capture the power law degree distribution that is observed in many real world networks. In the standard BA construction, a small random seed network with $m_0$ nodes is specified, then each new node is attached to $m\leq m_0$ existing nodes, with preference to high degree nodes. Thus the minimum degree of any node is $m$ and the mean degree is $k = 2m$. We modify the preferential attachment protocol in order to gain independent control of the minimum degree and the mean degree. We modify the BA algorithm by specifying that each new node connects to $i$ existing nodes, where $1\leq m_1\leq i\leq m_2 \leq m_0$ is drawn uniformly randomly for each new node node ($m_1$ and $m_2$ are parameters that replace the parameter $m$). This means that instead of a minimum degree equal to $m$, the minimum degree is $m_1$, which can be $1$ if desired, and the mean degree is $k = m_1+m_2$. This modification preserves the scale-free property of the networks generated, as demonstrated by the power law degree distribution $n\sim d^a$ shown in Fig.~\ref{fig:modified_BA_degree_distribution}, while at the same time allowing small minimum degrees and independent control of the minimum and mean degrees.

\section{Collective coordinate model reduction}  \label{sec:standard_CC}

The general idea of the collective coordinate approach is to specify a low-dimensional, judiciously chosen, ansatz manifold. The dynamics within the ansatz manifold is then determined via orthogonal projection of the full system \cite{Gottwald15}. Thus the approach can be thought of as a Galerkin approximation.

We now detail how to perform the collective coordinate approximation. For $K>K_g$, such that all oscillators are synchronized, it is natural to consider an ansatz consisting of a single synchronized cluster. A suitable ansatz manifold is found by linearizing the full Kuramoto model (\ref{eq:full_KM}) about $\phi_i = \phi_j$ for all $i$ and $j$, i.e.,
\begin{equation} \label{eq:linear_KM}
\dot{\bm \phi} \approx \bm\omega - \frac{K}{N} L \, \bm \phi ,
\end{equation}
where 
\begin{equation} \nonumber
L = D-A
\end{equation} 
is the graph Laplacian with $D$ being the diagonal degree matrix \cite{HancockGottwald18}. The synchronized (stationary) solution of (\ref{eq:linear_KM}) is given by
 \begin{equation} \label{eq:linear_KM_solution}
\bm \phi \approx  \frac{N}{K} L^+ \, \bm \omega ,
\end{equation}
where $L^+$ denotes the pseudo-inverse of $L$. The vector $\hat{\bm \phi} = L^+ \, \bm \omega$ provides a basis for the ansatz manifold, so that the full ansatz manifold is given by
\begin{equation} \label{eq:single_cluster_ansatz_function}
\bm \phi \approx \alpha(t) \hat{\bm \phi},
\end{equation}
where $\alpha(t)$ is called the collective coordinate. Note that the factor $N/K$ in (\ref{eq:linear_KM_solution}) is absorbed by $\alpha$. The dynamics of $\alpha$ is found by minimizing the error associated with restricting the phase space to the ansatz manifold. The error vector is given by
\begin{equation} \label{eq:single_cluster_error}
\mathcal{E}_i = \dot \alpha \hat{\phi}_i - \omega_i - \frac{K}{N}\sum_{j=1}^N A_{ij} \sin\left(\alpha(\hat\phi_j -\hat\phi_i)\right), 
\end{equation}
for $i=1,\dots,N$, and is minimized provided that it is orthogonal to the space spanned by the ansatz manifold given by (\ref{eq:single_cluster_ansatz_function}), i.e., $\langle \hat{\bm\phi}, \bm{\mathcal{E}} \rangle = 0$, where $\langle \bm u ,\bm v \rangle = \bm u^T \bm v$ denotes the Euclidean inner product. Orthogonality yields the evolution equation
\begin{equation} \label{eq:single_cluster_evo_equation_micro}
\dot{\alpha} = \frac{\langle \hat{\bm\phi},\bm\omega\rangle}{\langle \hat{\bm\phi}, \hat{\bm\phi}\rangle} + \frac{K}{N \langle \hat{\bm\phi}, \hat{\bm\phi}\rangle} \sum_{i,j=1}^N \hat{\phi}_i A_{ij} \sin\left( \alpha ( \hat\phi_j - \hat\phi_i) \right),
\end{equation}
which is a one-dimensional ordinary differential equation for the collective coordinate $\alpha$. This approach has been successfully used to approximate the collective dynamics of the Kuramoto model \cite{Gottwald15, SmithGottwald20}, including generalizations that describe the inter- and intra-cluster dynamics that results from topological clustering \cite{HancockGottwald18} or frequency clustering \cite{SmithGottwald19}. The approach has also been successfully applied to the stochastic Kuramoto model \cite{Gottwald17} and the Kuramoto-Sakaguchi model \cite{YueEtAl20} which includes phase frustration.

Synchronized states of the full Kuramoto model (\ref{eq:full_KM}) are approximated via the collective coordinate framework by stable stationary solutions of the evolution equation (\ref{eq:single_cluster_evo_equation_micro}). As well as approximating synchronized solutions, the evolution equation (\ref{eq:single_cluster_evo_equation_micro}) also describes the approximate dynamics of the system, encoding information such as the relaxation rate of perturbations away from synchronized solutions and bifurcation structure when stability is lost \cite{Gottwald15, Gottwald17, HancockGottwald18, SmithGottwald19, SmithGottwald20, YueEtAl20}. The collective coordinate method is designed to capture the dynamics of synchronized states. At the bifurcation point $K=K_g$, from global synchronization to partial synchronization, the dynamics of individual oscillators, or a small group of oscillators, that break off from the synchronized cluster becomes important. For instance, we will show that the standard collective coordinate approach results in large errors ($25-35\%$) in the estimation of $K_g$ for sparse networks. In Section~\ref{sec:single_cluster_simplified} we will show how this error can be corrected.

The critical coupling strength $K_g$ corresponding to global synchronization can be approximated via the collective coordinate approach through a combination of two criteria \cite{HancockGottwald18}. 
\begin{defn} \label{defn:Kg_criteria}
Under the collective coordinate description, $K_g$ is the smallest value of $K$ such that 
\begin{enumerate}[(i)]
\item A stable stationary solution $\alpha^*$ to (\ref{eq:single_cluster_evo_equation_micro}) exists and,
\item The approximation $\bm\phi^* = \alpha^* \hat{\bm \phi}$ of the synchronized state is stable in the full Kuramoto model.
\end{enumerate}
\end{defn}
The second criterion reflects the fact that even if the solution $\alpha^*$ is stable under the reduced dynamics (\ref{eq:single_cluster_evo_equation_micro}), the state $\bm\phi^* =\alpha^* \hat{\bm \phi}$ may not be stable in the full system (\ref{eq:full_KM}), as quantified by the eigenvalues of the Jacobian of (\ref{eq:full_KM}) evaluated at $\bm\phi^*$, i.e.,
\begin{equation} \label{eq:Llin}
\left(L_\text{lin}\right)_{ij} = \frac{K}{N} \begin{cases} 
-\sum_{k=1}^N A_{ik} \cos\left(\alpha^* (\hat\phi_k - \hat\phi_i) \right) &,\, i=j  \\ 
A_{ij} \cos \left(\alpha^* (\hat\phi_j - \hat\phi_i) \right) &,\, i\neq j
\end{cases}
\end{equation}
The matrix $L_\text{lin}$ always has one zero eigenvalue, and if all the other eigenvalues are negative then the state $\bm\phi^*$ is stable in the full system and the second criterion is met. If a single eigenvalue is positive then the state $\bm\phi^* $ is unstable in the full system and the second criterion is not met. This second stability criterion detects if the full system has already undergone a saddle-node bifurcation, such that the synchronized state loses stability. The node that is first to break off from the synchronized can be identified using this criterion as the dominant term in the eigenvector corresponding to the unstable eigenvalue \cite{HancockGottwald18}.

\begin{figure*}[tbp]
\centering
\includegraphics[width=0.8\textwidth]{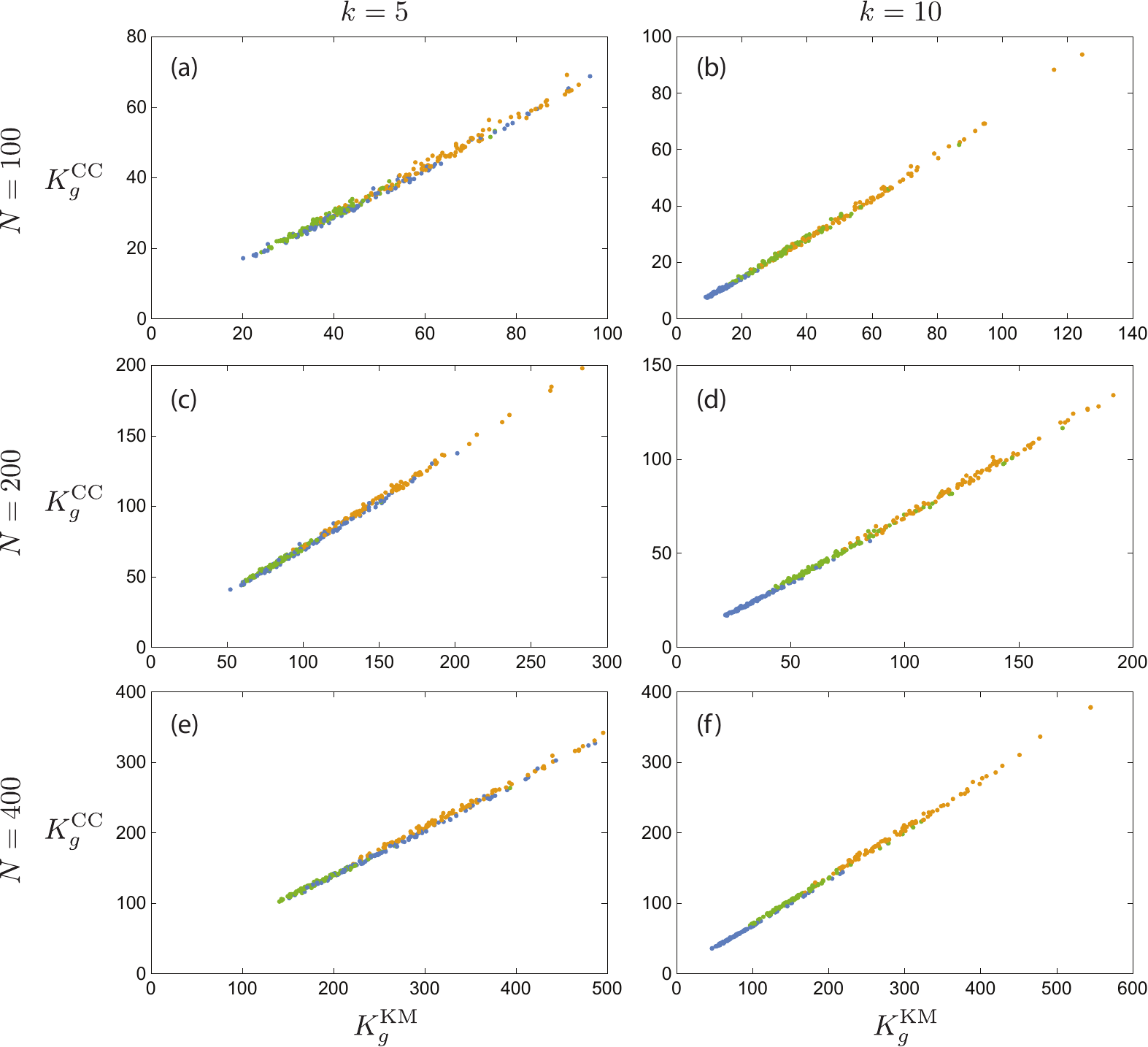}
\caption{
Critical coupling strength $K_g$ in sparse networks obtained from the full Kuramoto model (\ref{eq:full_KM}), denoted $\KgKM$, and obtained via the collective coordinate approach (\ref{eq:single_cluster_evo_equation_micro}), denoted $\KgCC$ for different numbers of oscillators $N$, mean degree $k$, and network structures. Results are shown for 100 realizations each of Erd\H os-R\'enyi graphs (blue) and modified Barab\'asi-Albert graphs with minimum degree $m_1 =1$ (orange) and $m_1 =2$ (green). In all plots both the network structure $A$ and the natural frequencies $\omega_i$ are randomly realized. Note that there is significant overlap between the data-sets, especially for $k=5$. (a,b)~$N=100$, (c,d)~$N=200$, (e,f)~$N=400$. (a,c,e)~$k=5$, (b,d,f)~$k=10$.
}
\label{fig:Kg_CCvsKM_sparse}
\end{figure*}

We denote the critical coupling strength corresponding to the bifurcation from global synchronization to partial synchronization obtained from the full model (\ref{eq:full_KM}) by $\KgKM$, and the critical coupling strength obtained from the collective coordinate method by $\KgCC$. We compute $\KgKM$ and $\KgCC$ for many realizations of random network topologies and random natural frequencies. For the full model, $\KgKM$ is computed numerically by finding the smallest value of $K$ such that (\ref{eq:full_KM}) admits a stable stationary solution \footnote{For each value of $K$, stationary states are found by solving $\dot \phi_i = 0$ for $i=1,\dots,N$ in the full Kuramoto model (\ref{eq:full_KM}) using the multidimensional Newton root finding method. If a stationary state is found, then its stability is checked by determining the eigenvalues of the Jacobian. We perform bisection in $K$ until $K_g$ is found within a tolerance of $10^{-4}$.}. In all cases the natural frequencies are drawn from a Gaussian distribution with variance $\sigma^2 = 0.1$ and are uniformly shifted so that $\sum \omega_i =0$. The random network topologies consist of ER networks and modified BA networks of various sizes $N$ and mean degree $k$. 
When we generate networks for our numerical simulations, we do not enforce the structure of Fig.~\ref{fig:degree_one_network_schematic} on the networks. Some, but not all, networks will have at least one degree 1 node, but, as we will show, the results also apply to sparse networks with minimum degree greater than 1. In the case of BA networks we explicitly control the minimum degree through the parameter $m_1$.
Fig.~\ref{fig:Kg_CCvsKM_sparse} shows $\KgKM$ vs $\KgCC$ for sparse networks, such that $k \ll N$. For all values of $k$ and $N$, and both ER and BA networks, we observe an approximately linear relationship between $\KgKM$ and $\KgCC$. For the sparse BA networks the minimum degree is controlled, and, as expected, networks with a minimum degree $m_1 = 1$ (orange) have higher values of $K_g$ compared to those with $m_1 = 2$ (green). For the ER networks (blue), the minimum degree is not controlled. For $k=5$ most ER networks have minimum degree equal to one (66\% for $N=100$, 78\% for $N=200$ and 89\% for $N=400$) and so there is significant overlap between the data-sets for ER and BA networks. For $k=10$ the ER networks typically have minimum degree greater than two, and so the $K_g$ values are smaller than for the BA networks. Of particular importance, the relationship between $\KgKM$ and $\KgCC$ does not depend strongly on either the network construction (ER vs BA) or the minimum degree of the network. Note, for instance, that all of the ER networks generated with $N=100$ and $k=10$ have minimum degree greater than one, but they follow the same trend as the BA networks with minimum degree equal to one. In Section~\ref{sec:single_cluster_simplified} we derive an analytic expression for this relationship between $\KgKM$ and $\KgCC$, assuming that the network has the structure in Fig.~\ref{fig:degree_one_network_schematic}.

\begin{figure}[tbp]
\centering
\includegraphics[width=0.9\columnwidth]{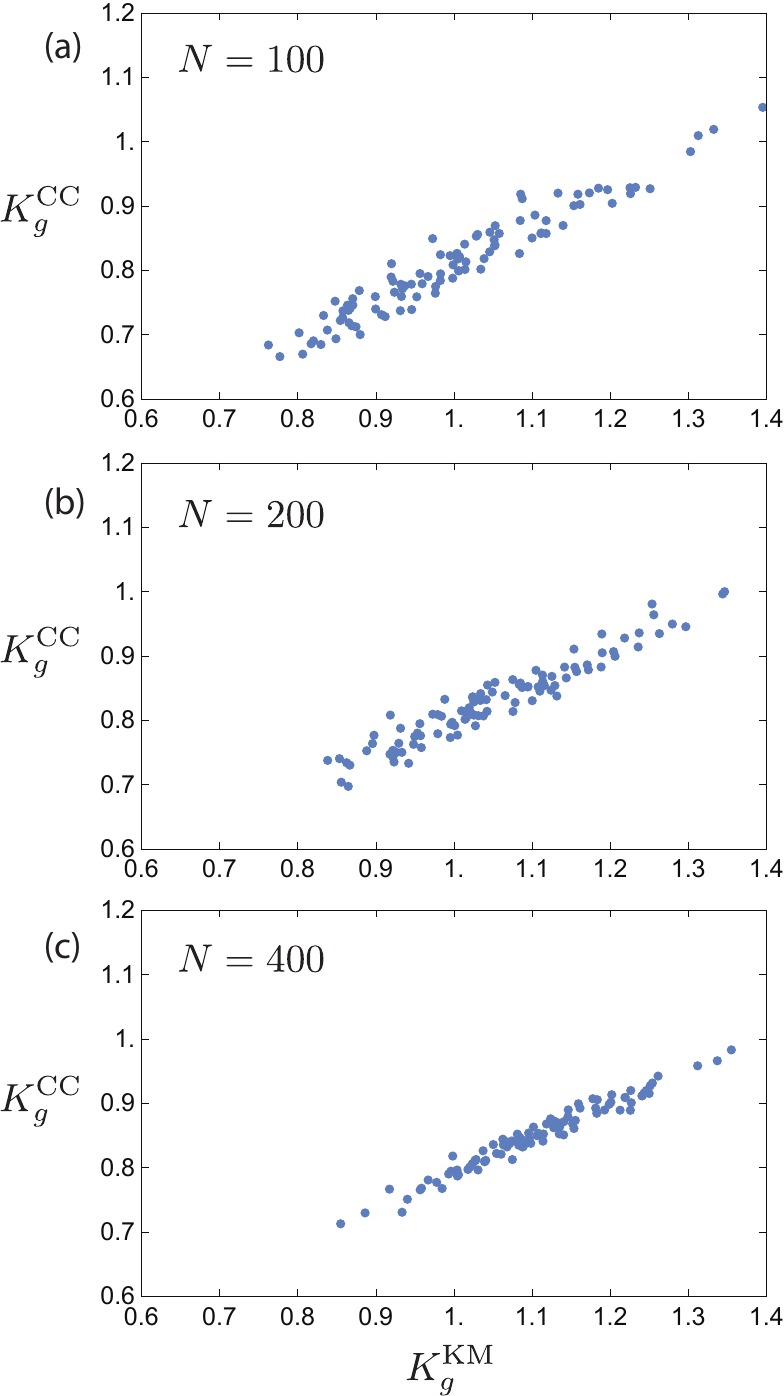}
\caption{
Critical coupling strength $K_g$ in dense networks obtained from the full Kuramoto model (\ref{eq:full_KM}), denoted $\KgKM$, and obtained via the collective coordinate approach (\ref{eq:single_cluster_evo_equation_micro}), denoted $\KgCC$ for different numbers of oscillators $N$ for Erd\H os-R\'enyi networks with $p=0.95$. Results are shown for 100 realizations of the network structure $A_{ij}$ and the natural frequencies $\omega_i$. (a)~$N=100$, (b)~$N=200$, (c)~$N=400$.
}
\label{fig:Kg_CCvsKM_dense}
\end{figure}

For dense networks, there is still a positive correlation between $\KgCC$ and $\KgKM$, as shown in Fig.~\ref{fig:Kg_CCvsKM_dense} for ER networks with coupling probability $p=0.95$. However, there is much more deviation compared to the results for sparse networks shown in Fig.~\ref{fig:Kg_CCvsKM_sparse}, especially for smaller networks ($N=100$) in which finite size effects have a greater impact. For the dense networks used in Fig.~\ref{fig:Kg_CCvsKM_dense}, the error in approximating the true vale $\KgKM$ by the collective coordinate approach $\KgCC$ is in the range $10-25\%$, which is significantly more accurate than for the sparse networks in Fig.~\ref{fig:Kg_CCvsKM_sparse} for which the error is $25-35\%$. Therefore, we will focus our analysis on sparse networks.

\section{Mesoscopic ansatz function for a single collective coordinate} \label{sec:single_cluster_simplified}

We now describe a method that incorporates known ensemble information about the network and natural frequencies, together with knowledge of which node is most susceptible to desynchronization, to significantly reduce the computational complexity of the evolution equation (\ref{eq:single_cluster_evo_equation_micro}) while preserving the most important information. We assume as in the standard collective coordinate approach in Section~\ref{sec:standard_CC} a single collective coordinate $\alpha(t)$ with linearized ansatz manifold (\ref{eq:single_cluster_ansatz_function}) describing the main synchronized cluster. We shall use here the particular structure of the network (cf. Fig.~\ref{fig:degree_one_network_schematic}) to modify the ansatz manifold (\ref{eq:single_cluster_ansatz_function}) by taking ensemble averages over realizations of the network structure $A$ and natural frequencies $\bm{\omega}$. This significantly reduces the complexity and we will see that the reduced description also determines the approximately linear relationship between $\KgKM$ and $\KgCC$, which can be used to correct approximations $\KgCC$ obtained from the full description (\ref{eq:single_cluster_evo_equation_micro}). The reduction proceeds as an averaging over many realizations of the random graph and random frequencies under the assumption that the networks have the structure shown in Fig.~\ref{fig:degree_one_network_schematic}, where $|\omega_1|$ is such that node 1 is the first to desynchronize. Averaging over many realizations of graphs with mean degree $k$, the mean connectivity between any two nodes, excluding node 1, is equal to $p = k/(N-1)$. Here we will denote the average of a variable $x$ over many realizations of random networks and random frequencies by $\langle{x}\rangle$. The averaged structure of the network has adjacency matrix $\langle A\rangle$ such that 
\begin{align}
\langle A\rangle_{12} &= \langle A\rangle_{21} = 1,  \nonumber \\
\langle A\rangle_{1j} &= \langle A\rangle_{j1}  = 0, \quad j>2,  \label{eq:averaged_network_structure} \\
\langle A\rangle_{ii} & = 0, \quad 1\leq i \leq N, \nonumber \\ 
\langle A\rangle_{ij} &= p, \quad \text{otherwise.} \nonumber
\end{align}
Note that the ensemble averaged adjacency matrix (\ref{eq:averaged_network_structure}) is identical for both ER and BA networks, and so some information about the degree distribution, such as the presence of hubs, is lost in the ensemble average. We will see that our averaging method, leading to the adjacency matrix (\ref{eq:averaged_network_structure}) is highly accurate for both ER and BA networks, suggesting that the information lost does not contribute significantly to the outcome. We remark that in order to retain information about eventual hubs, one may compute the ensemble average by retaining the degrees of each node by prescribing the degrees $k_i$ of the nodes, following the specified degree distribution, and averaging over network structures with those nodal degrees \cite{RodriguesEtAl16, PeronEtAl19}. 
Having obtained the averaged adjacency matrix (\ref{eq:averaged_network_structure}), it can then be shown that the pseudo-inverse of the averaged graph Laplacian $\langle L\rangle$ is given by $\langle L\rangle^+ = B /(pN^2)$ where
\begin{align}
B_{11} &= N-2 + p(N-1)^2, \nonumber \\
B_{1j} &= B_{j1} = N-2 - p(N-1), \quad  j>1,  \nonumber \\
B_{22} &= N-2 + p,  \nonumber \\
B_{2j} &= B_{j2} = p-2, \quad j>2,  \nonumber  \\
B_{jj} &= p + N + \frac{2}{N-1}, \quad 3\leq j\leq N,   \nonumber  \\
B_{ij} &= B_{ji} = p - \frac{N-2}{N-1}, \quad 3\leq i < j \leq N.   \nonumber 
\end{align}

The natural frequencies are written as 
\begin{equation} \nonumber
\bm\omega = (\omega_1, \omega_2, \Omega_3 + \xi_3, \dots \Omega_3 + \xi_N)
\end{equation}
where $\Omega_3 = \frac{1}{N-2} \sum_{j=3}^N \omega_j$ is the mean frequency of the remainder of the network and $\xi_3,\dots,\xi_N$ denote deviations from the mean. Since the total mean frequency may always be set to zero through a suitable change of coordinates, we have that
\begin{equation} \label{eq:mean_zero_frequency}
\Omega_3 = -\frac{\omega_1 + \omega_2}{N-2}.
\end{equation}
For any function $f$, averaging over frequency realizations yields the identity
\begin{equation}  \label{eq:averaging_general_functions}
\left\langle \frac{1}{N-2}\sum_{j=3}^N f(\xi_j) \right\rangle = \int_{-\infty}^\infty f(\xi) (g(\xi) - \mu) d\xi
\end{equation}
where $g(\xi)$ is the probability density function from which the natural frequencies $\omega_i$ are drawn, and $\mu$ is its mean. Here the frequencies are drawn from a Gaussian distribution $g(\omega)$ with mean $\mu = 0$ and variance $\sigma^2$. In particular, for $f(\xi) =\xi $ and $f(\xi) = \xi^2$ (\ref{eq:averaging_general_functions}) yields the averaged mean and averaged variance
\begin{align}
\left\langle \frac{1}{N-2}\sum_{j=3}^N \xi_j \right\rangle &= 0,   \label{eq:averaged_mean}   \\
\left\langle \frac{1}{N-2}\sum_{j=3}^N \xi_j^2 \right\rangle &= \sigma^2.
\end{align}
Similarly, we obtain
\begin{align}
\left\langle \frac{1}{N-2}\sum_{j=3}^N \sin(\xi_j) \right\rangle &= 0,    \\
\left\langle \frac{1}{N-2}\sum_{j=3}^N \cos(\xi_j) \right\rangle &= e^{-\sigma^2/2}.  \label{eq:averaged_cosine}
\end{align}

Having computed the averaged graph Laplacian $\langle L \rangle$ and its pseudo-inverse $\langle L \rangle ^+$, we may compute the averaged collective coordinate ansatz function $\hat{ \bm{\phi}} = \langle L \rangle ^+ \bm \omega$ for a given realization of frequencies $\bm \omega = (\omega_1, \omega_2, \Omega_3 + \xi_3,\dots,\Omega_3 + \xi_N)$. After some algebra, it can be shown that
\begin{align} 
\hat\phi_1 & = \frac{1}{pN}\left( \left( p(N-1) +1 \right)\omega_1 + \omega_2\right) \nonumber \\
\hat\phi_2 &= -\frac{1}{pN}\left( p \,\omega_1 + (N-2)\Omega_3 \right)  \label{eq:single_cluster_ansatz_function_averaged} \\
\hat\phi_j &= \frac{1}{pN}\left( -p\,\omega_1 + 2\Omega_3 \right) + \frac{1}{p(N-1)} \xi_j, \quad j=3,\dots,N.  \nonumber
\end{align}
 Substituting this ansatz function and the averaged adjacency matrix $\langle A \rangle$ (\ref{eq:averaged_network_structure}) into the collective coordinate evolution equation (\ref{eq:single_cluster_evo_equation_micro}), and averaging over frequency realizations using (\ref{eq:averaged_mean})-(\ref{eq:averaged_cosine}), yields the simplified evolution equation for the collective coordinate
\begin{equation} \label{eq:single_cluster_evo_eq_reduced}
\dot{\alpha} = \frac{a - b \frac{K}{N}}{c},
\end{equation}
where
\begin{align}
a =&  \,\omega_1^2 + \frac{N-2}{p} \Omega_3^2 + \frac{(N-2)\sigma^2}{(N-1)p},  \nonumber \\
b =& \frac{\alpha (N-2)\sigma^2 E }{(N-1)^2 p} \left((N-2)E + \cos\left(\frac{\alpha \Omega_3}{p} \right) \right) +  \nonumber \\
 & \quad \omega_1 \sin(\alpha \omega_1) + (N-2) E \Omega_3 \sin\left(\frac{\alpha \Omega_3}{p}\right), \label{eq:single_cluster_evo_eq_reduced_parameters}   \\
c = & \frac{(N-1) \omega_1^2}{N} - \frac{2(N-2) \omega_1 \Omega_3}{Np} + \nonumber  \\
 & \quad \frac{N-2}{p^2}\left( \frac{\sigma^2}{(N-1)^2} + \frac{2 \Omega_3^2}{N} \right), \nonumber 
\end{align}
with
\begin{equation}
E = \exp\left(-\frac{1}{2}\left(\frac{\alpha \sigma}{(N-1)p}\right)^2\right). \nonumber 
\end{equation}
Here we have used the zero mean frequency condition (\ref{eq:mean_zero_frequency}) to eliminate $\omega_2$ as a parameter ($\omega_2 = -\omega_1 - (N-1) \Omega_3$). The reduced evolution equation (\ref{eq:single_cluster_evo_eq_reduced}) has far fewer parameters than the standard collective coordinate evolution equation (\ref{eq:single_cluster_evo_equation_micro}), six parameters compared to $N^2 + N$ parameters. In addition, there are no summations in the reduced equation (\ref{eq:single_cluster_evo_eq_reduced}) while there is a double summation over $N$ terms in the standard equation (\ref{eq:single_cluster_evo_equation_micro}). Therefore, the reduced equation is vastly less computationally complex ($\mathcal{O}(1)$) compared to the standard equation ($\mathcal{O}(N^2)$).

\begin{figure}[tbp]
\centering
\includegraphics[width=0.9\columnwidth]{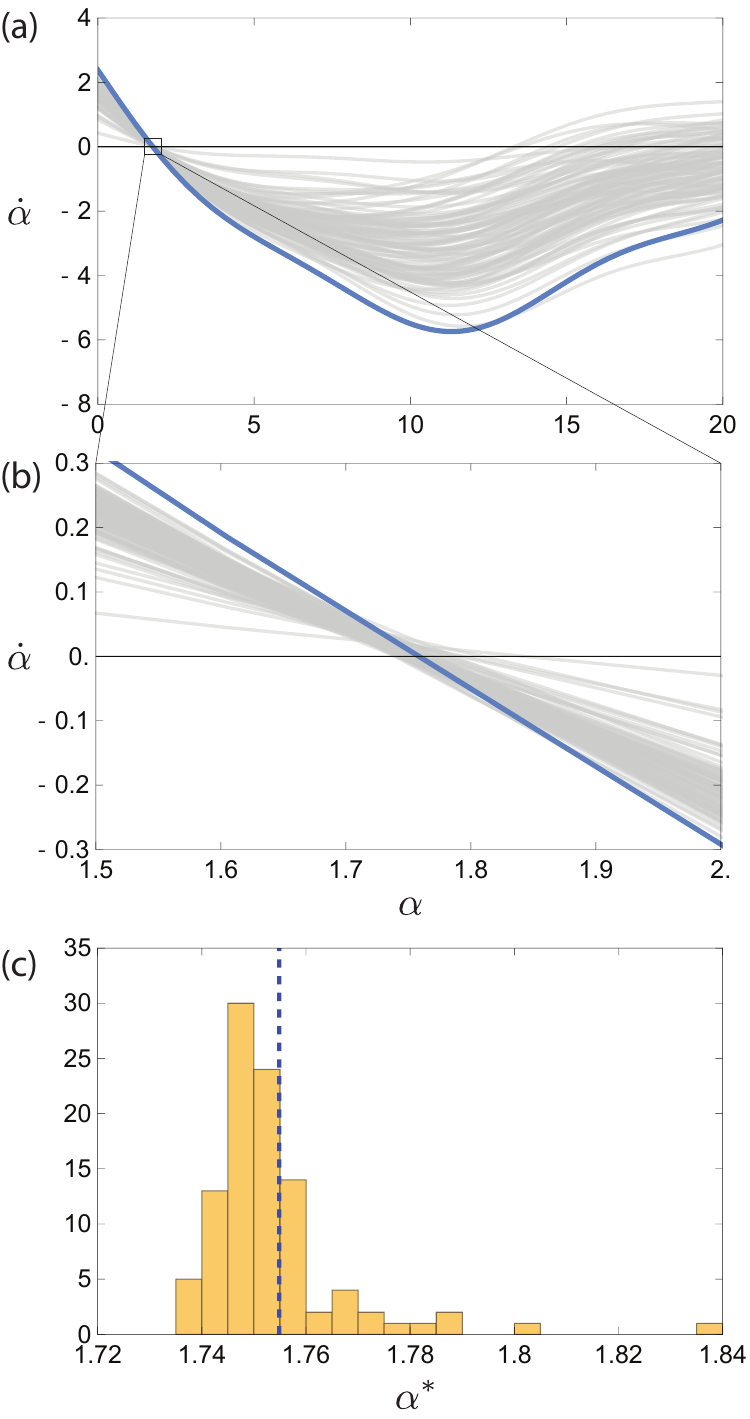}
\caption{
(a,b)~100 realizations of the collective coordinate evolution equation (\ref{eq:single_cluster_evo_equation_micro}) (light gray) for coupling strength $K=60$ and Erd\H os-R\'enyi networks with $N=100$ and $k=5$ such that node 1 has degree one and is connected to node 2, and natural frequencies such that $\omega_1 = -0.677$, $\omega_2 = 0.0434$ and $\omega_3,\dots,\omega_N$ are drawn randomly from a Gaussian distribution with variance $\sigma^2 = 0.1$ and such that $\sum_{i=1}^N \omega_i = 0$. The reduced collective coordinate evolution equation (\ref{eq:single_cluster_evo_eq_reduced}) is shown in dark blue. (b)~Subregion of (a) near the $\dot{\alpha} = 0$ intercept. (c)~Histogram of the stationary points $\alpha^*$ such that $\dot{\alpha}(\alpha^*) = 0$ for the 100 realizations in (a,b). The intercept point $\alpha^* = 1.754$ of the reduced equation (\ref{eq:single_cluster_evo_eq_reduced}) is shown as the vertical dashed line.
}
\label{fig:alpha-dot_averaged}
\end{figure}

The result of the averaging process is demonstrated in Fig.~\ref{fig:alpha-dot_averaged}(a,b), where the gray curves show the right hand side of the standard evolution equation (\ref{eq:single_cluster_evo_equation_micro}) for 100 realizations of random Erd\H os-R\'enyi graphs and random frequencies, keeping the coupling strength $K=60$ fixed. The random graphs have $N=100$ and $k=5$, and all are constructed to have the structure shown in Fig.~\ref{fig:degree_one_network_schematic}.
The random frequencies $\omega_i$ are drawn such that $\omega_1$ and $\omega_2$ are kept fixed, while the remainder of the frequencies are drawn randomly from the Gaussian distribution $g(\omega)$ with variance $\sigma^2=0.1$. The frequencies $\omega_3, \dots, \omega_N$ are then uniformly shifted to ensure the zero mean frequency condition (\ref{eq:mean_zero_frequency}) is satisfied, i.e., so that $\Omega_3 = -(\omega_1 + \omega_2)/(N-2)$. We present here results for $\omega_1 = -0.677$ and $\omega_2 = 0.0434$, implying $\Omega_3 = 0.00646$.
The thick blue curves in Fig.~\ref{fig:alpha-dot_averaged}(a,b) show the right hand side of the averaged equation (\ref{eq:single_cluster_evo_eq_reduced}). We observe that while the averaged curve is not the average of the realizations for all values of $\alpha$, it is close to the average of the realizations near the intercept $\dot\alpha = 0$ (Fig.~\ref{fig:alpha-dot_averaged}(b)). Considering the histogram of such stationary points, i.e., $\alpha^*$ such that $\dot\alpha(\alpha^*) = 0$, which is shown in Fig.~\ref{fig:alpha-dot_averaged}(c) for the 100 realizations, we observe that their mean, which is $\overline{\alpha^*} = 1.753$, agrees with the stationary solution of the averaged evolution equation (\ref{eq:single_cluster_evo_eq_reduced}) (vertical dashed line at $\alpha^* = 1.754$). It is arguably this stationary solution that is the most important aspect of the collective coordinate framework, as it determines the approximate synchronized state, and is critical in determining $K_g$ through the collective coordinate method. The small range of stationary point values produced by all the realizations of the network structure and natural frequencies indicates that the macroscopic dynamics of the full Kuramoto model (\ref{eq:full_KM}) is insensitive to changes in the network structure and the natural frequencies of nodes $3,\dots,N$, as long as the statistical properties ($k$, $\Omega_3$ and $\sigma^2$) are unchanged. This is further evidence that the averaging approach is justified. 
We remark that changing the value of the coupling strength $K$ does not qualitatively change the results of Fig.~\ref{fig:alpha-dot_averaged}, the stationary points of the individual realizations (\ref{eq:single_cluster_evo_equation_micro}) still agree strongly with the stationary point of the averaged equation (\ref{eq:single_cluster_evo_eq_reduced}). The only difference is that as $K$ increases, the minima of all curves decrease, and the stationary points shift toward $\alpha = 0$.

The simplification achieved by averaging allows to find the relationship between $\KgCC$ and $\KgKM$ for sparse networks as observed in Fig.~\ref{fig:Kg_CCvsKM_sparse}, and, hence, correct the estimate $\KgCC$ without needing to know which node will desynchronize first. In Section~\ref{sec:KM_model} it was shown that for networks with structure as in Fig.~\ref{fig:degree_one_network_schematic} such that node 1 is the first to desynchronize we have $\KgKM = |\omega_1| N$. To determine $\KgCC$ we must consider the two criteria in Definition~\ref{defn:Kg_criteria}. We consider criterion~(ii) of Definition~\ref{defn:Kg_criteria}, stability of the approximate solution $\hat{\bm\phi}^* = \alpha^* \hat{\bm \phi}$ in the full Kuramoto model (\ref{eq:full_KM}), where $\alpha^*$ is the stationary point of the collective coordinate evolution equation (\ref{eq:single_cluster_evo_eq_reduced}), because this criterion is stronger than criterion~(i). Stability in the full Kuramoto model (\ref{eq:full_KM}) is determined by the eigenvalues of $L_\text{lin}$ (\ref{eq:Llin}), or, equivalently, the eigenvalues of the Jacobian $J$ of the phase-difference system, given by (\ref{eq:differenced_Jacobian}) for networks with the structure in Fig.~\ref{fig:degree_one_network_schematic}. Following the same analysis as in Section~\ref{sec:KM_model}, at $\Phi_1 \coloneqq \phi_2 - \phi_1  = \pi/2$, $J$ has a zero eigenvalue and $\det J$ changes sign crossing the hyperplane $\Phi_1 = \pi/2$. Therefore, stability of $\hat{\bm\phi}^* = \alpha^* \hat{\bm \phi}$ is lost exactly when $\hat{\bm\Phi}^* $, defined componentwise by $\hat{\Phi}^*_i =\hat{\phi}^*_{i+1} - \hat{\phi}^*_{i}$, coincides with the hyperplane $\Phi_1 = \pi/2$, such that $\alpha^* (\hat \phi_2 - \hat \phi_1) = \pi/2$. As such, $\KgCC$ is the value of $K$ such that
\begin{equation} \label{eq:single_cluster_CC_condition}
\dot{\alpha}|_{\alpha =  (\pi/2)/ (\hat\phi_2 - \hat\phi_1)} = 0.
\end{equation}
The condition (\ref{eq:single_cluster_CC_condition}) is valid for both the standard collective coordinate evolution equation (\ref{eq:single_cluster_evo_equation_micro}), provided the network has the structure in Fig.~\ref{fig:degree_one_network_schematic}, and for the reduced collective coordinate evolution equation (\ref{eq:single_cluster_evo_eq_reduced}). Whereas this expression is intractable for the standard collective coordinate approach, requiring all the microscopic parameters $A_{ij}$ and $\omega_i$, it is readily available for the mesoscopic reduction (\ref{eq:single_cluster_evo_eq_reduced}). For the mesoscopic description the solution is
\begin{equation} \label{eq:single_cluster_reduced_CC_solution}
\KgCC(\omega_1,\Omega_3,N,p,\sigma^2) = \frac{aN}{b} \bigg|_{\alpha =  (\pi/2)/ (\hat\phi_2 - \hat\phi_1)}
\end{equation}
such that $a$ and $b$ are as in (\ref{eq:single_cluster_evo_eq_reduced_parameters}). We explicitly state all the parameters that $\KgCC$ depends on via $a$ and $b$. This is a computationally simple solution, although writing the full expression is tedious.

\begin{figure*}[tbp]
\centering
\includegraphics[width=0.8\textwidth]{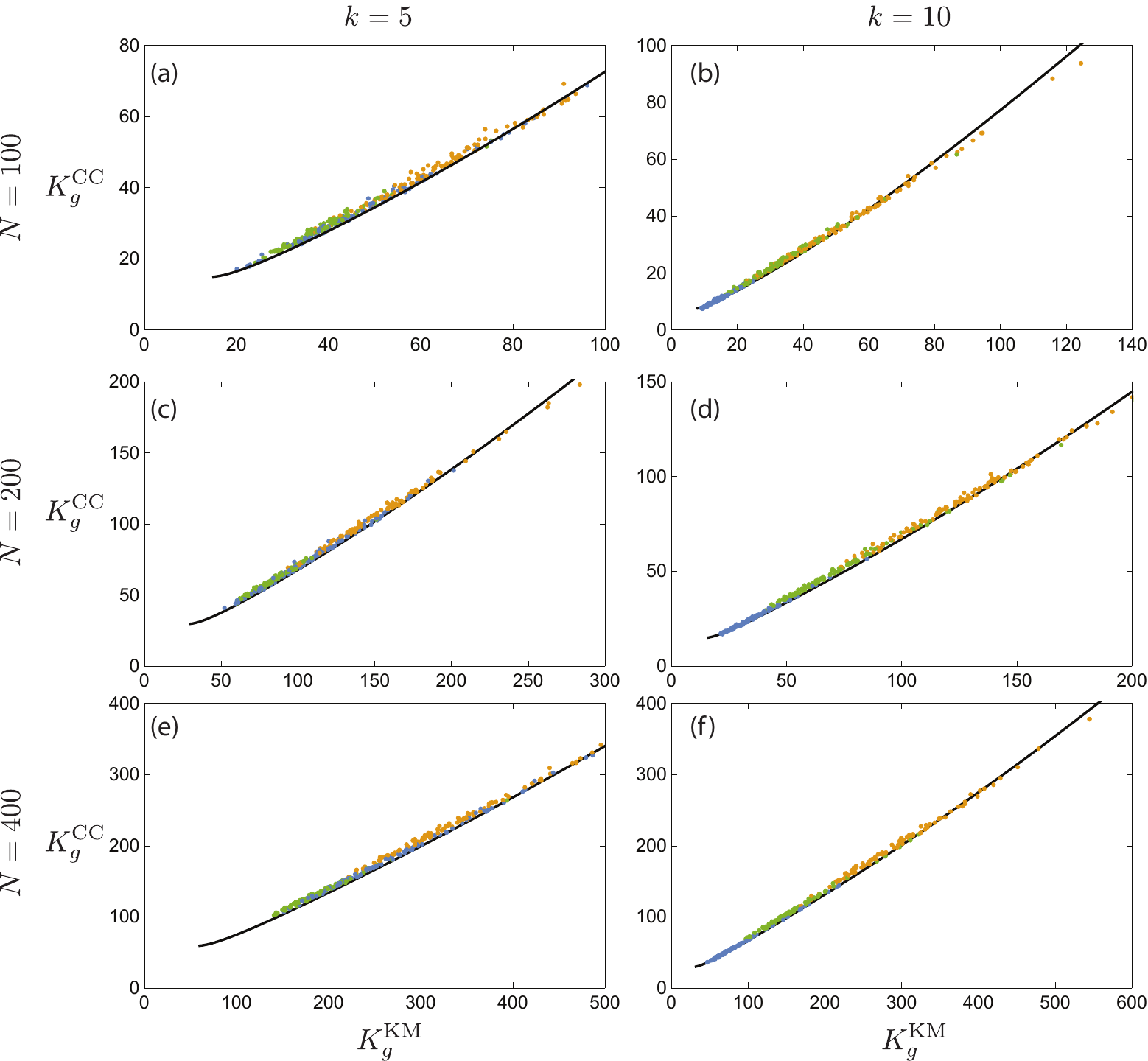}
\caption{
The relationship $\KgCC(\KgKM;N,p,\sigma^2)$ described by (\ref{eq:single_cluster_reduced_CC_relationship_2}) (continuous black curve) for sparse networks with the same values of $N$ and $k = (N-1)p$ as in Fig.~\ref{fig:Kg_CCvsKM_sparse}, together with the data in Fig.~\ref{fig:Kg_CCvsKM_sparse} obtained from random realizations of network topologies (both Erd\H os-R\'enyi topologies (blue points) and modified Barab\'asi-Albert topologies (orange points for $m_1 =1$ and green points for $m_1 = 2$)) and natural frequencies. (a,b)~$N=100$, (c,d)~$N=200$, (e,f)~$N=400$. (a,c,e)~$k=5$, (b,d,f)~$k=10$.
}
\label{fig:Kg_CCvsKM_analytic}
\end{figure*}

Since $\KgKM = |\omega_1| N$, we have $\omega_1 = \KgKM / N$ (assuming, without loss of generality, that $\omega_1>0$). In addition, from the zero mean frequency condition (\ref{eq:mean_zero_frequency}) we have
\begin{equation} \nonumber
 \Omega_3 = -\frac{\omega_1 + \omega_2}{N-2} \approx -\frac{\omega_1}{N-2} = -\frac{\KgKM/N}{N-2}.
\end{equation} 
Substituting these expressions for $\omega_1$ and $\Omega_3$ into (\ref{eq:single_cluster_reduced_CC_solution}) yields the relationship between $\KgCC$ and $\KgKM$ as follows
\begin{equation} \label{eq:single_cluster_reduced_CC_relationship_2}
\KgCC(\KgKM;N,p,\sigma^2) = \frac{aN}{b} \bigg|_{\begin{scriptsize}
\begin{array}{l} \alpha =  \frac{\pi/2}{\hat\phi_2 - \hat\phi_1},\, \omega_1 = \frac{\KgKM}{N},\,   \\ \Omega_3 = -\frac{\KgKM/N}{N-2} \end{array}
\end{scriptsize}},
\end{equation}
which depends only on the ensemble parameters $N$, $p$ and $\sigma^2$. This relationship is shown in Fig.~\ref{fig:Kg_CCvsKM_analytic} for the same values of $N$ and $k = (N-1)p$ as the first two columns of Fig.~\ref{fig:Kg_CCvsKM_sparse}, overlaying the data from Fig.~\ref{fig:Kg_CCvsKM_sparse} for random realizations of network structures and frequencies. It is clear that the simple relationship (\ref{eq:single_cluster_reduced_CC_relationship_2}) accurately captures the approximately linear relationship observed from the random realizations in Fig.~\ref{fig:Kg_CCvsKM_sparse}.

In our elimination of $\Omega_3$ to obtain (\ref{eq:single_cluster_reduced_CC_relationship_2}) through the substitution $\Omega_3 = -\frac{\KgKM/N}{N-2}$ we ignore variations in $\Omega_3$ that result from $\omega_2 \neq 0$. This is justified because variation in $\Omega_3$ has a negligible effect on the relationship between $\KgCC$ and $\KgKM$ (results not shown).

\begin{figure}[tbp]
\centering
\includegraphics[width=0.9\columnwidth]{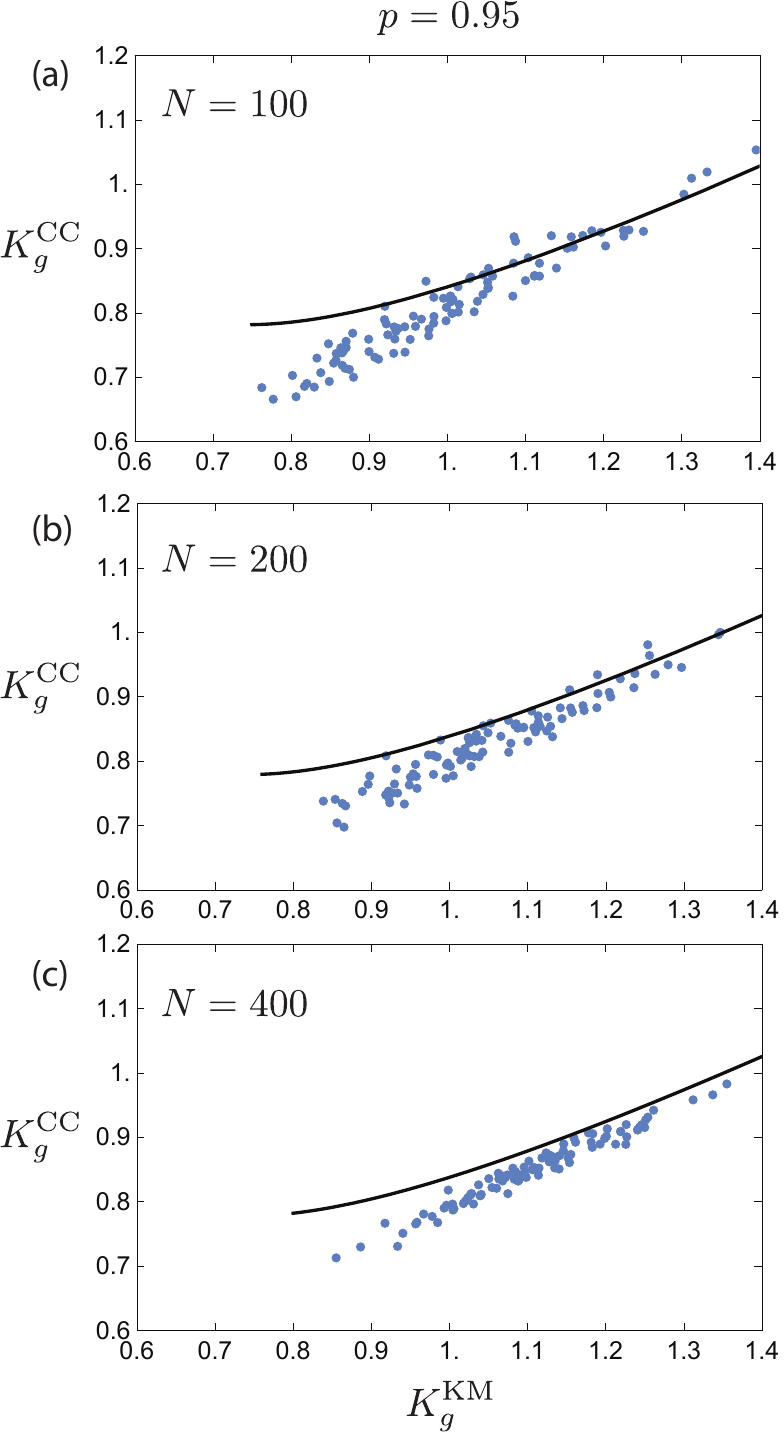}
\caption{
The relationship $\KgCC(\KgKM;N,p,\sigma^2)$ described by (\ref{eq:single_cluster_reduced_CC_relationship_2}) (continuous black curve) for dense networks with the same values of $N$ and $p = 0.95$ as in Fig.~\ref{fig:Kg_CCvsKM_dense}, together with the data in Fig.~\ref{fig:Kg_CCvsKM_dense} obtained from random realizations of Erd\H os-R\'enyi network topologies and natural frequencies.  (a)~$N=100$, (b)~$N=200$, (c)~$N=400$.
}
\label{fig:Kg_CCvsKM_analytic_dense}
\end{figure}

For the dense networks considered in Fig.~\ref{fig:Kg_CCvsKM_dense} with $p=0.95$ we find that the relationship (\ref{eq:single_cluster_reduced_CC_relationship_2}) does not accurately capture the trend between $\KgCC$ and $\KgKM$. This is shown in Fig.~\ref{fig:Kg_CCvsKM_analytic_dense}. It is not surprising that the description (\ref{eq:single_cluster_reduced_CC_relationship_2}) is poor in these cases, because their dense network structures are far from the assumed structure shown in Fig.~\ref{fig:degree_one_network_schematic} which has a degree one node. The probability of a degree one node existing scales as $(1-p)^N$, which is extremely small for $p\approx 1$ and large $N$. Similarly, the probability of low degree nodes with degree greater than one are very small.

\subsection*{Discussion}

We propose the relationship (\ref{eq:single_cluster_reduced_CC_relationship_2}) as a means to correct $\KgCC$ obtained from the standard collective coordinate description (\ref{eq:single_cluster_evo_equation_micro}). That is, given $\KgCC$ obtained from (\ref{eq:single_cluster_evo_equation_micro}) and only knowledge of $N$, $p$ and $\sigma^2$, we can invert (numerically) (\ref{eq:single_cluster_reduced_CC_relationship_2}) to obtain an improved approximation to $\KgKM$. Using (\ref{eq:single_cluster_reduced_CC_relationship_2}) to correct the standard collective coordinate approach has the advantage that it requires no knowledge of the microscopic dynamics of the full system. In particular, one does not need any prior knowledge of which node is the first to desynchronize.

In addition, the relationship between $\KgCC$ and $\KgKM$ is not only valid for networks of the form Fig.~\ref{fig:degree_one_network_schematic}, but also for sparse networks in which the node that breaks off has degree greater than 1. This is demonstrated in Fig.~\ref{fig:Kg_CCvsKM_analytic} by the BA networks with minimum degree $m_1 = 2$ and by the ER networks, many of which have minimum degree greater than one when $N=100$. For networks with minimum degree greater than 1, and for which there are many low degree nodes, determining which node will be the first to desynchronize is more complex, requiring computation of the eigenvalues and eigenvectors of the Jacobian for coupling strengths close to the critical coupling strength $K_g$. On the other hand, the standard collective coordinate method avoids this computation, as it does not require knowledge of the most susceptible node. So, from a practical standpoint, it is beneficial to use the standard collective coordinate approach and the correction given by (\ref{eq:single_cluster_reduced_CC_relationship_2}).

\section{Mesoscopic ansatz function for four collective coordinates} \label{sec:three_cluster}

In the previous section we obtained a correction to the critical coupling strength for the standard collective coordinate approach. However, the temporal dynamics of the full system near the bifurcation at $K_g$ are still not well described by a single collective coordinate. To accurately describe the dynamics we consider a new mesoscopic ansatz function with four collective coordinates. This derives from the collective coordinate approach to treating multiple interacting synchronized clusters \cite{Gottwald15, SmithGottwald19}, as occurs for multimodal frequency distributions or topological clustering. The difference here is that two of the ``clusters'' consist of individual nodes, nodes $1$ and $2$ in a network of the form Fig.~\ref{fig:degree_one_network_schematic}. The third cluster consists of the remainder of the nodes (nodes $3,\dots,N$). Our description here is mesoscopic in the sense that it captures the microscopic details of nodes $1$ and $2$, but only the macroscopic detail of nodes $3,\dots,N$.


The mesoscopic collective coordinate ansatz function is 
\begin{equation} \label{eq:three_cluster_ansatz_function}
\bm{\phi} \approx \hat{\bm \phi} = f_1(t) \bm{\chi}_1
+ f_2(t) \bm{\chi}_2
+ f_3(t) \bm{\chi}_3
+
\alpha_3(t) \bm{\mathcal{S}}_3,
\end{equation}
where $\bm{\chi}_i$ is the indicator vector for each cluster, i.e., $\bm{\chi}_1 = (1,0,0,\dots)$, $\bm{\chi}_2 = (0,1,0,\dots)$ and $\bm{\chi}_3 = (0,0,1,1,\dots)$, $f_i(t)$ describes the mean phase of the $i$-th cluster, $\alpha_3(t)$ is a dilation parameter which controls the shape of the synchronized third cluster together with the shape vector $\bm{\mathcal{S}}_3 = (0,0,\hat{\bm \theta}_3)$. Ordinarily, there would be shape vectors and dilation parameters for all the clusters, but here clusters 1 and 2 consist of single nodes and so do not require a description of the cluster shape. The ansatz (\ref{eq:three_cluster_ansatz_function}) reduces the dynamics from $N$ variables down to four collective coordinates. 

The shape function $\hat{\bm \theta}_3$ for the third cluster is derived by considering the third cluster in isolation (i.e., ignoring nodes 1 and 2) and solving the resultant linearized system (as for the single cluster case (\ref{eq:linear_KM})-(\ref{eq:linear_KM_solution})), yielding
\begin{equation}
\hat{\bm \theta}_3 = L_3^+ \bm{\omega}_3,
\end{equation}
where 
\begin{equation} \nonumber
L_3 = D_3 - A_3
\end{equation}
is the $(N-2)\times(N-2)$ graph Laplacian of the subgraph obtained by removing nodes 1 and 2 from the graph $A$, and $\bm{\omega}_3 = (\omega_3,\dots,\omega_N)$. The ansatz manifold $\hat{\bm \phi}$, onto which the dynamics of the full system is projected, is spanned by the four vectors, $\bm{\chi}_{1,2,3}$ and $\bm{\mathcal{S}}_3$, forming an orthogonal basis. Orthogonality of $\bm{\chi}_3$ and $\bm{\mathcal{S}}_3$ results from the fact that $L_3$ is a graph Laplacian and therefore has a zero eigenvalue with corresponding eigenvector $(1,1,\dots)$. 

The error vector associated with restricting the phase space of the full Kuramoto model (\ref{eq:full_KM}) to the ansatz manifold is given by
\begin{equation}
\mathcal{E}_i =  \sum_{j=1}^3 \dot f_j \chi_j^{(i)} + \dot\alpha_3 \mathcal{S}_3^{(i)} - \omega_i  - \frac{K}{N}\sum_{j=1}^N A_{ij} \sin\left( \hat\phi_j - \hat\phi_i \right),
\end{equation}
for $i=1,\dots,N$. This error is minimized when it is orthogonal to the ansatz manifold, i.e., when $\langle \mathcal{E}, \bm{\chi}_{1,2,3}\rangle = 0$ and $\langle \mathcal{E}, \bm{\mathcal{S}}_3 \rangle = 0$. These orthogonality conditions yield the evolution equations for the collective coordinates with
\begin{align}
\dot{f}_1 &= \omega_1 + \frac{K}{N} \sin(f_2 - f_1), \label{eq:three_cluster_CC_evo_eqs_agnostic_1}  \\
\dot{f}_2 &= \omega_2 + \frac{K}{N} \left( \sin(f_1 - f_2) + \sum_{j=3}^N A_{2j} \sin(\hat{\phi}_j - f_2) \right),  \\
\dot{f}_3 &= \Omega_3 + \frac{K}{N(N-2)} \sum_{k=3}^N \sum_{j=2}^N A_{jk} \sin(\hat\phi_j - \hat\phi_k),  \\
\dot\alpha_3 &= \frac{\hat{\bm \theta}_3^T \bm\omega_3}{\hat{\bm \theta}_3^T \hat{\bm \theta}_3} + \frac{K}{N\hat{\bm \theta}_3^T \hat{\bm \theta}_3} \sum_{k=3}^N \sum_{j=2}^N \mathcal{S}_3^{(k)} A_{jk} \sin(\hat\phi_j - \hat\phi_k), \label{eq:three_cluster_CC_evo_eqs_agnostic_4}
\end{align}
where $\Omega_3 = \frac{1}{N-2} \sum_{j=3}^N \omega_j$ as before. Unlike the collective coordinate description with a single collective coordinate (\ref{eq:single_cluster_evo_equation_micro}), the evolution equations (\ref{eq:three_cluster_CC_evo_eqs_agnostic_1})-(\ref{eq:three_cluster_CC_evo_eqs_agnostic_4}) require knowledge of the oscillator most susceptible to desynchronization in order to distinguish nodes 1 and 2. Furthermore, the three cluster description requires the network to have the form Fig.~\ref{fig:degree_one_network_schematic}, which is not required for the standard collective coordinate description. We assume here that we have prior knowledge of the node that is first to desynchronize. In Section~\ref{sec:standard_CC} we explained how to obtain the critical node through the standard collective coordinate framework using the eigenvectors of $L_\text{lin}$ (\ref{eq:Llin}).

Following a similar procedure to Section~\ref{sec:single_cluster_simplified}, a significant simplification to the evolution equations (\ref{eq:three_cluster_CC_evo_eqs_agnostic_1})-(\ref{eq:three_cluster_CC_evo_eqs_agnostic_4}) is possible by averaging the network structure and natural frequencies over many realizations. The averaged network structure $A_{ij}$ is as in (\ref{eq:averaged_network_structure}), and frequencies are written again in the form $\bm\omega = (\omega_1, \omega_2, \Omega_3 + \xi_3, \dots, \Omega_3 + \xi_N)$. After averaging the network structure, the subnetwork for nodes $3,\dots,N$ has adjacency matrix $\langle A_3 \rangle = p(\bm{1}_{N-2}^T \bm{1}_{N-2} - I_{N-2})$, i.e., $\langle A_3 \rangle$ represents an all-to-all connected graph with uniform connection weight $p$. It can be shown that $\langle L_3 \rangle^+ = \frac{1}{(N-2)^2p^2} \langle L_3 \rangle$, which results in
\begin{equation}
\hat{\bm \theta}_3 = \langle L_3 \rangle^+ \bm\omega_3 = \frac{1}{(N-2)p} (\xi_3, \dots, \xi_N)^T.
\end{equation}
Substituting $\hat{\bm \theta}_3$ into the evolution equations (\ref{eq:three_cluster_CC_evo_eqs_agnostic_1})-(\ref{eq:three_cluster_CC_evo_eqs_agnostic_4}) and averaging over frequency realizations using the identities (\ref{eq:averaged_mean}-(\ref{eq:averaged_cosine})) we obtain the reduced set of equations
\begin{align}
\dot{f}_1 &= \omega_1 + \frac{K}{N} \sin(f_2 - f_1), \label{eq:three_cluster_CC_evo_eqs_simplified_1}  \\
\dot{f}_2 &= \omega_2 + \frac{K}{N} \left( \sin(f_1 - f_2) + \phantom{e^{-\frac{\beta_3^2\sigma^2}{2}}} \right. \nonumber \\
& \qquad \qquad \qquad  \left. (N-2)p \sin(f_3 - f_2) e^{-\frac{\beta_3^2\sigma^2}{2}} \right),  \\
\dot{f}_3 &= \Omega_3 + \frac{K}{N} p \sin(f_2-f_3)e^{-\frac{\beta_3^2\sigma^2}{2}},  \\
\dot\beta_3 &= 1 \! - \! \frac{K}{N} p \beta_3 e^{-\frac{\beta_3^2\sigma^2}{2}} \left( \cos(f_2 - f_3) + (N-2) e^{-\frac{\beta_3^2\sigma^2}{2}} \right),  \label{eq:three_cluster_CC_evo_eqs_simplified_4}
\end{align}
where we have rescaled the dilation parameter $\alpha_3$ by letting $\beta_3(t) = \frac{1}{(N-2)p}\alpha_3(t)$. This simplified description provides a significant reduction in computational complexity, compared to both the full Kuramoto model (\ref{eq:full_KM}) and the full collective coordinate description (\ref{eq:three_cluster_CC_evo_eqs_agnostic_1})-(\ref{eq:three_cluster_CC_evo_eqs_agnostic_4}). 

Note that equations (\ref{eq:three_cluster_CC_evo_eqs_agnostic_1}) and (\ref{eq:three_cluster_CC_evo_eqs_simplified_1}) for $\dot f_1$ are identical, and have the same form as the equation for $\dot\phi_1$ in the full Kuramoto model (\ref{eq:full_KM}). Therefore, by the same reasoning as in Section~\ref{sec:KM_model} we obtain $\KgCC = \KgKM = |\omega_1| N$, with a saddle-node bifurcation occurring in the collective coordinate dynamics (\ref{eq:three_cluster_CC_evo_eqs_agnostic_1})-(\ref{eq:three_cluster_CC_evo_eqs_agnostic_4}) at $K = \KgCC$.

\begin{figure*}[tbp]
\centering
\includegraphics[width=0.8\textwidth]{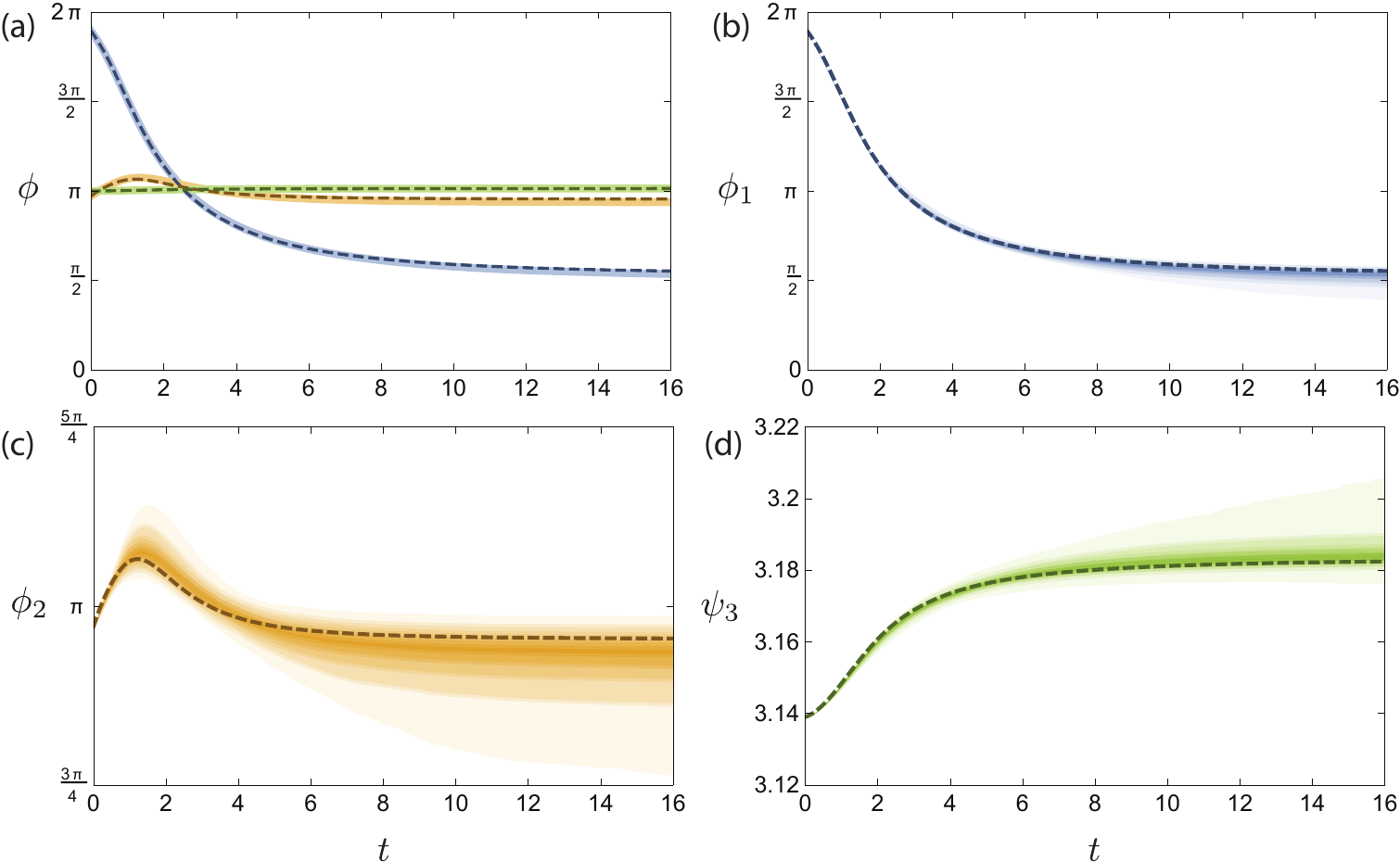}
\caption{
Trajectories of the collective coordinate phases $f_1$ (dashed blue), $f_2$ (dashed orange) and $f_3$ (dashed green) with dynamics governed by the reduced mesoscopic dynamics (\ref{eq:three_cluster_CC_evo_eqs_simplified_1})-(\ref{eq:three_cluster_CC_evo_eqs_simplified_4}) and statistical properties of $1000$ trajectories of the full Kuramoto model (\ref{eq:full_KM}) for $N=100$, $p=0.05$, $\omega_1 = -0.677$, $\omega_2 = 0.0434$, $\Omega_3 = 0.00646$, and $\sigma^2 = 0.1$, for globally synchronized dynamics with coupling strength $K = 70> K_g = 67.7$. For the full Kuramoto model the frequencies $\bm \omega$ and network topologies $A$ are randomized as described in the text. (a)~Trajectories of the collective coordinates $f_1$, $f_2$ and $f_3$ together with the medians of the trajectories of the full Kuramoto model for each of $\phi_1$ (thick solid blue), $\phi_2$ (thick solid orange) and the mean phase $\psi_3$ of the third cluster defined by (\ref{eq:psi_3}) (thick solid green). (b-d)~Trajectories of the collective coordinates $f_1$, $f_2$ and $f_3$, respectively, each combined with shaded regions corresponding to quantile ranges of trajectories of the full model. For example, the darkest shaded band of each color corresponds to the quantile range $45-55\%$, and each successive decrease in darkness corresponds to a quantile range containing $10\%$ of trajectories.
}
\label{fig:three_cluster_K>Kg}
\end{figure*}

To show that the reduced mesoscopic equations (\ref{eq:three_cluster_CC_evo_eqs_simplified_1})-(\ref{eq:three_cluster_CC_evo_eqs_simplified_4}) accurately capture the dynamics of the full Kuramoto model (\ref{eq:full_KM}), we compare trajectories of the full system with trajectories of the reduced system. We will consider both coupling strengths $K>K_g$, such that there is convergence to a stationary state, and $K<K_g$, such that node 1 desynchronizes resulting in non-stationary dynamics. For the full Kuramoto model (\ref{eq:full_KM}), in each case we consider trajectories for $1000$ realizations of randomly drawn frequencies and randomly generated networks with $N=100$ oscillators. As in Section ~\ref{sec:single_cluster_simplified}, for the frequencies, we fix the values $\omega_1 = -0.677$, $\omega_2 = 0.0434$, $\Omega_3 = 0.00646$, and $\sigma^2 = 0.1$, and randomly draw the remaining frequencies $\omega_3,\dots,\omega_N$ from a normal distribution with variance $\sigma^2$, and enforce that the mean frequency $\frac{1}{N-2}\sum_{j=3}^N \omega_j = \Omega_3$. For the network structure, we generate networks with the form in Fig.~\ref{fig:degree_one_network_schematic}, such that the subnetwork for nodes $2,\dots,N$ is an Erd\H os-R\'enyi network with $p=0.05$. For these systems, the critical coupling strength is $K_g = N |\omega_1| = 67.7$. Fig.~\ref{fig:three_cluster_K>Kg} shows the trajectories of the phases for $K=70 > K_g$, with all trajectories having the same initial condition. The trajectories of the reduced mesoscopic system (\ref{eq:three_cluster_CC_evo_eqs_simplified_1})-(\ref{eq:three_cluster_CC_evo_eqs_simplified_4}) are shown by dashed curves, such that $\phi_1$ (blue), $\phi_2$ (orange), and $\psi_3$ (green) correspond to the collective coordinates $f_1$, $f_2$, and $f_3$, respectively. For the full model, we show statistical information from the $1000$ realizations, and note that $\psi_3$ is used to denote the mean phase of nodes $3,\dots,N$, i.e.,
\begin{equation} \label{eq:psi_3}
\psi_3 = \text{arg} \left( \frac{1}{N-2} \sum_{j=3}^N e^{i\phi_j} \right).
\end{equation}
Fig.~\ref{fig:three_cluster_K>Kg}(a) shows the median curves (solid thick curves) for each of $\phi_1$ (blue), $\phi_2$ (orange) and $\psi_3$ (green), such that we compute the median of the $1000$ trajectories at each time instant. We see very good agreement between the median trajectories of the full Kuramoto model and the trajectories of the reduced system (\ref{eq:three_cluster_CC_evo_eqs_simplified_1})-(\ref{eq:three_cluster_CC_evo_eqs_simplified_4}). Figs.~\ref{fig:three_cluster_K>Kg}(b)-\ref{fig:three_cluster_K>Kg}(d) show more detailed statistical information for the trajectories of $\phi_1$, $\phi_2$ and $\psi_3$, respectively. The shading in each figure corresponds to quantile ranges, such that the darkest shading corresponds to the quantile range $0.45-0.55$ (containing the median), and each successive decrease in the darkness corresponds to a quantile range that also contains $10\%$ of the trajectories. We observe that most trajectories are contained within a narrow range, and this narrow range agrees very well with the reduced system (\ref{eq:three_cluster_CC_evo_eqs_simplified_1})-(\ref{eq:three_cluster_CC_evo_eqs_simplified_4}). We checked that outlying trajectories correspond to pathological realizations of the frequencies and network, such as having node 2 connected to only a single node in the main synchronized cluster, resulting in nodes 1 and 2 desynchronizing as a pair.

\begin{figure*}[tbp]
\centering
\includegraphics[width=0.8\textwidth]{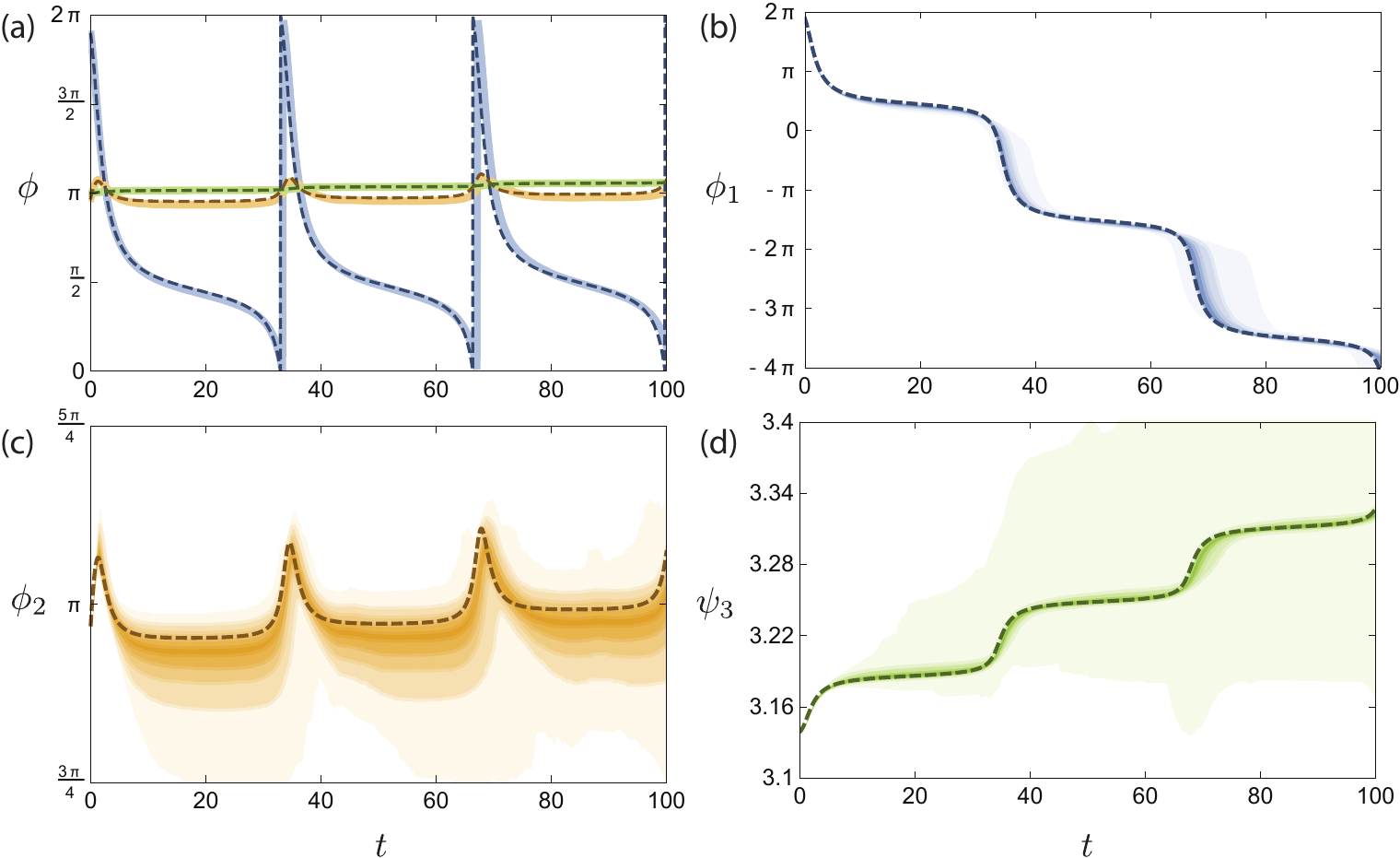}
\caption{
As for Fig.~\ref{fig:three_cluster_K>Kg}, except for non-synchronized dynamics at $K = 65 < K_g$. (a)~Trajectories of the collective coordinates $f_1$, $f_2$ and $f_3$ together with the medians of the trajectories of the full Kuramoto model for each of $\phi_1$ (thick solid blue), $\phi_2$ (thick solid orange) and the mean phase $\psi_3$ of the third cluster defined by (\ref{eq:psi_3}) (thick solid green). (b-d)~Trajectories of the collective coordinates $f_1$, $f_2$ and $f_3$, respectively, each combined with shaded regions corresponding to quantile ranges of trajectories of the full model. For example, the darkest shaded band of each color corresponds to the quantile range $45-55\%$, and each successive decrease in darkness corresponds to a quantile range containing $10\%$ of trajectories.
}
\label{fig:three_cluster_K<Kg}
\end{figure*}

For the desynchronized state with $K = 65 < K_g$, Fig.~\ref{fig:three_cluster_K<Kg} shows the corresponding trajectories and statistical information, as for Fig.~\ref{fig:three_cluster_K>Kg}. The same random frequencies and network structures are used, and the same initial condition for all trajectories is used. In this case the dynamics is non-stationary, with oscillator 1 having desynchronized from the rest of the oscillators. We again observe very good agreement between the median of the trajectories of the full Kuramoto model (thick solid curves in Fig.~\ref{fig:three_cluster_K<Kg}(a)) and the trajectories of the reduced mesoscopic system (\ref{eq:three_cluster_CC_evo_eqs_simplified_1})-(\ref{eq:three_cluster_CC_evo_eqs_simplified_4}) (dashed curves). Furthermore, considering the quantile ranges in Figs.~\ref{fig:three_cluster_K<Kg}(b)-\ref{fig:three_cluster_K<Kg}(d), most trajectories of the full Kuramoto model (\ref{eq:full_KM}) fall within a narrow range, which contains the trajectories of the reduced mesoscopic system (\ref{eq:three_cluster_CC_evo_eqs_simplified_1})-(\ref{eq:three_cluster_CC_evo_eqs_simplified_4}). The results for both $K>K_g$ and $K<K_g$ show that the greatly simplified mesoscopic model (\ref{eq:three_cluster_CC_evo_eqs_simplified_1})-(\ref{eq:three_cluster_CC_evo_eqs_simplified_4}), which incorporates only the ensemble statistics of the frequencies and network structure for nodes $3,\dots,N$, is able to accurately reproduce the dynamics of the full Kuramoto model in most cases. This highlights that the specific details of the collectively organised main cluster is unimportant to the resultant dynamics.

The reduced mesoscopic collective coordinate system (\ref{eq:three_cluster_CC_evo_eqs_simplified_1})-(\ref{eq:three_cluster_CC_evo_eqs_simplified_4}) can be further simplified using a time-scale splitting between the phase dynamics of $f_{1,2,3}$ and the order parameter $r_3(t)$ of the third cluster (nodes $3,\dots,N$), similar to the time-scale splitting derived previously \cite{SmithGottwald19}. The collective coordinate approximation yields an order parameter $r_3$ of the third cluster given by
\begin{equation} \nonumber
r_3 = \left| \frac{1}{N-2} \sum_{j=3}^N e^{i \alpha_3 \hat\Phi_3^{(j)}} \right| = \left| \frac{1}{N-2} \sum_{j=3}^N e^{i \beta_3 \xi_j} \right|.
\end{equation}
Averaging over frequency realizations yields
\begin{equation} \label{eq:r3_of_b3}
\langle r_3 \rangle = e^{-\frac{\beta_3^2 \sigma^2}{2}}.
\end{equation}
Using (\ref{eq:r3_of_b3}) we can express the mesoscopic evolution equations (\ref{eq:three_cluster_CC_evo_eqs_simplified_1})-(\ref{eq:three_cluster_CC_evo_eqs_simplified_4}) in terms of $f_{1,2,3}$ and $r_3$ only, i.e.,
\begin{align}
\dot{f}_1 &= \omega_1 + \frac{K}{N} \sin(f_2 - f_1), \label{eq:three_cluster_CC_evo_eqs_r3_1}  \\
\dot{f}_2 &= \omega_2 + \frac{K}{N} \left( \sin(f_1 - f_2) + (N-2)p \sin(f_3 - f_2) r_3 \right),  \\
\dot{f}_3 &= \Omega_3 + \frac{K}{N} p \sin(f_2-f_3)r_3,  \\
\dot r_3 &= - r_3 \sqrt{\log r_3^{-2}} \left( \sigma - \frac{K}{N} p r_3 \sqrt{\log r_3^{-2}} \times   \right.  \nonumber  \\
& \left. \qquad \qquad \qquad \quad \phantom{\frac{K}{N}} \left( \cos(f_2 - f_3) + (N-2) r_3 \right) \right). \label{eq:three_cluster_CC_evo_eqs_r3_4}
\end{align}
These equations have the advantage that the physical meaning of $r_3$ is clearer than $\beta_3$. Furthermore, for coupling strengths close to, or greater than, $K_g$, the third cluster remains synchronized, and so $r_3(t) = 1 - \epsilon(t)$ with $0<\epsilon\ll 1$. Expanding the evolution equations (\ref{eq:three_cluster_CC_evo_eqs_r3_1})-(\ref{eq:three_cluster_CC_evo_eqs_r3_4}) around $\epsilon = 0$ to leading order yields
\begin{align}
\dot{f}_1 &= \omega_1 + \frac{K}{N} \sin(f_2 - f_1), \label{eq:three_cluster_CC_evo_eqs_eps_1}  \\
\dot{f}_2 &= \omega_2 + \frac{K}{N} \left( \sin(f_1 - f_2) + (N-2)p \sin(f_3 - f_2) \right),  \\
\dot{f}_3 &= \Omega_3 + \frac{K}{N} p \sin(f_2-f_3),  \label{eq:three_cluster_CC_evo_eqs_eps_3}  \\
\dot r_3 &= - \sqrt{2} \sigma \epsilon^{\frac{1}{2}} +\frac{2Kp}{N} \left(\cos(f_2-f_3) + N-2 \right) \epsilon, \label{eq:three_cluster_CC_evo_eqs_eps_4}
\end{align}
which demonstrates a time-scale splitting between the slow order parameter $r_3$ and the fast phases $f_{1,2,3}$. On the fast time scale, the dynamics is purely phase dynamics which are decoupled from the order parameter $r_3$. In the limit $\epsilon \to 0$ we obtain
\begin{align}
\dot{F}_1 &= \omega_2 - \omega_1 + \frac{K}{N} \left( - 2 \sin F_1)+  (N-2)p \sin F_2 \right),   \\
\dot{F}_2 &= \Omega_3 - \omega_2 + \frac{K}{N} \left(  \sin F_1 - (N-1) p \sin F_2 \right),
\end{align}
where $F_i = f_{i+1}-f_i$. This reduced system has stationary points
\begin{equation}
\sin F_1 = -\frac{N}{K}\omega_1, \text{ and } \sin F_2 = \frac{N}{K} \frac{\Omega_3}{p}.
\end{equation}
Thus, for large coupling strengths $K$, there are four stationary points. Stability analysis reveals a pair of saddles, a stable node and an unstable node. Upon decreasing $K$, two saddle-node bifurcations occur simultaneously at $K=K_g = N |\omega_1|$, such that all stationary points vanish. The saddle-node corresponding to the coalescence of the stable node and saddle describes the bifurcation from global synchronization to partial synchronization. The unstable node describes an unstable stationary configuration of oscillators.

\subsection*{Discussion}

The reduced mesoscopic equations (\ref{eq:three_cluster_CC_evo_eqs_eps_1})-(\ref{eq:three_cluster_CC_evo_eqs_eps_4}) reveal a connection between the mesoscopic collective coordinate ansatz for one collective coordinate discussed in Section~\ref{sec:single_cluster_simplified} and the mesoscopic collective coordinate ansatz for four collective coordinates. In deriving the ansatz function for a single collective coordinate, we solve the linearized dynamics of the full Kuramoto model (\ref{eq:full_KM}), linearizing about $\phi_i - \phi_j = 0$. If instead we solve the linearized phase-only dynamics (\ref{eq:three_cluster_CC_evo_eqs_eps_1})-(\ref{eq:three_cluster_CC_evo_eqs_eps_3}), linearizing about $f_i - f_j =0$, we obtain the ansatz function
\begin{align}
\hat f_1 & = \frac{1}{pN}\left( \left( p(N-1) +1 \right)\omega_1 + \omega_2\right) \nonumber \\
\hat f_2 &= -\frac{1}{pN}\left( p \,\omega_1 + (N-2)\Omega_3 \right)  \nonumber \\
\hat f_3 &= \frac{1}{pN}(-p\omega_1 + 2\Omega_3),   \nonumber
\end{align}
which is identical to the ansatz function (\ref{eq:single_cluster_ansatz_function_averaged}) obtained for a single collective coordinate, except that the variations $\xi_j$ are not accounted for.

The accuracy of the reduction (\ref{eq:three_cluster_CC_evo_eqs_simplified_1})-(\ref{eq:three_cluster_CC_evo_eqs_simplified_4}), which condenses the parametric description of the majority of the system down to its ensemble statistics demonstrates that the microscopic parametric details of the bulk network are largely unimportant in describing the macroscopic dynamics of the system. The mesoscopic ansatz captures all the essential information.

For any specific realization of the network topology and natural frequencies, the original collective coordinate evolution equations (\ref{eq:three_cluster_CC_evo_eqs_agnostic_1})-(\ref{eq:three_cluster_CC_evo_eqs_agnostic_4}) with four collective coordinates are more accurate in describing the temporal phase dynamics of individual oscillators than the reduced mesoscopic equations (\ref{eq:three_cluster_CC_evo_eqs_simplified_1})-(\ref{eq:three_cluster_CC_evo_eqs_simplified_4}). This is not surprising because they incorporate the full microscopic parametric details of the system. However, this loss of accuracy of the reduced mesoscopic system is greatly offset by its analytical simplicity, its generality (applying to all realizations simultaneously), and its reduced computational complexity ($\mathcal{O}(N^2)$ for the original system (\ref{eq:three_cluster_CC_evo_eqs_agnostic_1})-(\ref{eq:three_cluster_CC_evo_eqs_agnostic_4}) and $\mathcal{O}(1)$ for the reduced mesoscopic system (\ref{eq:three_cluster_CC_evo_eqs_simplified_1})-(\ref{eq:three_cluster_CC_evo_eqs_simplified_4})). As noted previously, both collective coordinate methods obtain the exact value of $K_g$.

\section{Conclusions} \label{sec:conclusions}

We have introduced two mesoscopic collective coordinate reductions for the finite size Kuramoto model with sparse connectivity which result in significant reductions in complexity and which more accurately describe the dynamics of the full system for coupling strengths $K$ close to the critical coupling strength $K_g$ corresponding to global synchronization. It is typical of sparse networks that the first node to desynchronize is degree one (or low degree). We exploit this property by assuming a simplified, averaged, network structure, which captures the microscopic details of the important nodes: the one that first desynchronizes and the one it connects to, while averaging the collective behavior of the rest of the network. 

The mesoscopic reduction with one collective coordinate achieves an analytic expression that describes the observed relationship between the critical coupling strength obtained via the standard collective approach $\KgCC$ and the actual critical coupling strength $\KgKM$. This analytic expression depends only on the ensemble parameters of the full system, i.e., the number of oscillators $N$, the network mean degree $k$ and the variance of the natural frequencies $\sigma^2$. Using this relationship between $\KgCC$ and $\KgKM$, one can correct the estimate for $K_g$ obtained through the standard collective coordinate framework, which, while inaccurate for sparse networks, has the advantage that it does not require prior knowledge of which oscillator will be first to desynchronize. Furthermore, the relationship between $\KgCC$ and $\KgKM$ is valid for the more complex case where the node that first desynchronizes has degree greater than one. If there are many low degree nodes there is significant computational effort in determining which will desynchronize first. This problem can be avoided by employing the standard collective coordinate framework and then correcting the estimate for $K_g$.

The mesoscopic reduction with four collective coordinates uses two collective coordinates to capture the microscopic dynamics of the two most important oscillators, the one that first desynchronizes and the one it connects to, and two collective coordinates to describe the collective macroscopic dynamics of the remainder of the oscillators. Averaging over network configurations and natural frequency realizations yields a simplified description that we have shown accurately captures the dynamics of the full system in the more complex case of coupling strengths close to the critical coupling strength $K_g$. The simplified system, which reduces the microscopic details of the main synchronized cluster to its ensemble statistics, captures both convergence to a synchronized state for coupling strengths $K>K_g$ and non-stationary dynamics for coupling strengths $K<K_g$.

Both mesoscopic reductions highlight that for sparse networks the fine details of the network and frequencies for nodes that remain highly synchronized play only a small role in the resultant dynamics of the full system, since they can be replaced by ensemble statistical parameters with typically only a small loss in accuracy. The dynamics is dominated by the microscopic interactions of two critical nodes and the collective behavior of the remainder of the network.

Here we have focused on the bifurcation from global synchronization to partial synchronization at the critical coupling strength $K_g$. Our analysis can equally be applied to each successive bifurcation such that another node breaks off from the synchronized cluster. This can be performed by either ignoring the effects of nodes that have already desynchronized \cite{Gottwald15, HancockGottwald18, SmithGottwald20}, or by including the effect of non-synchronized oscillators through averaging \cite{YueEtAl20}. The latter approach is especially important if the dynamics of the full model includes a phase frustration parameter, as in the Kuramoto-Sakaguchi model \cite{SakaguchiKuramoto86}.

Our approach is designed to deal with finite size networks away from the thermodynamic limit. This allows for applications to real-world systems. In particular the issue of onset of global synchronization is important for power grids \cite{YangEtAl17}, which on short time scales can be described by Kuramoto-type models \cite{NishikawaMotter15}. Power outages can be caused by a loss of synchronization and identifying nodes which are likely to cause cascading failures is of utmost importance \cite{YangEtAl17,YangEtAl17b}. The detrimental role of degree 1 nodes (or so called dead ends) in the grid stability has been established for example within the Northern European power grid \cite{MenckEtAl14}. Furthermore, it was shown that the collective dynamical behavior in power grids is strongly influenced by finite size effects \cite{OlmiEtAl14,RodriguesEtAl16}. We hope that our computationally cheap method will be useful for studying and controlling large but finite real-world networks such as power grids.

\begin{acknowledgments}
We wish to acknowledge support from the Australian Research Council, Grant No. DP180101991.
\end{acknowledgments}

\section*{Data Availability}

The data that support the findings of this study are available from the corresponding author upon reasonable request.

\bibliography{sparse_networks.bib}

\begin{thebibliography}{32}%
\makeatletter
\providecommand \@ifxundefined [1]{%
 \@ifx{#1\undefined}
}%
\providecommand \@ifnum [1]{%
 \ifnum #1\expandafter \@firstoftwo
 \else \expandafter \@secondoftwo
 \fi
}%
\providecommand \@ifx [1]{%
 \ifx #1\expandafter \@firstoftwo
 \else \expandafter \@secondoftwo
 \fi
}%
\providecommand \natexlab [1]{#1}%
\providecommand \enquote  [1]{``#1''}%
\providecommand \bibnamefont  [1]{#1}%
\providecommand \bibfnamefont [1]{#1}%
\providecommand \citenamefont [1]{#1}%
\providecommand \href@noop [0]{\@secondoftwo}%
\providecommand \href [0]{\begingroup \@sanitize@url \@href}%
\providecommand \@href[1]{\@@startlink{#1}\@@href}%
\providecommand \@@href[1]{\endgroup#1\@@endlink}%
\providecommand \@sanitize@url [0]{\catcode `\\12\catcode `\$12\catcode
  `\&12\catcode `\#12\catcode `\^12\catcode `\_12\catcode `\%12\relax}%
\providecommand \@@startlink[1]{}%
\providecommand \@@endlink[0]{}%
\providecommand \url  [0]{\begingroup\@sanitize@url \@url }%
\providecommand \@url [1]{\endgroup\@href {#1}{\urlprefix }}%
\providecommand \urlprefix  [0]{URL }%
\providecommand \Eprint [0]{\href }%
\providecommand \doibase [0]{http://dx.doi.org/}%
\providecommand \selectlanguage [0]{\@gobble}%
\providecommand \bibinfo  [0]{\@secondoftwo}%
\providecommand \bibfield  [0]{\@secondoftwo}%
\providecommand \translation [1]{[#1]}%
\providecommand \BibitemOpen [0]{}%
\providecommand \bibitemStop [0]{}%
\providecommand \bibitemNoStop [0]{.\EOS\space}%
\providecommand \EOS [0]{\spacefactor3000\relax}%
\providecommand \BibitemShut  [1]{\csname bibitem#1\endcsname}%
\let\auto@bib@innerbib\@empty
\bibitem [{\citenamefont {Sheeba}, \citenamefont {Stefanovska},\ and\
  \citenamefont {McClintock}(2008)}]{SheebaEtAl08}%
  \BibitemOpen
  \bibfield  {author} {\bibinfo {author} {\bibfnamefont {J.~H.}\ \bibnamefont
  {Sheeba}}, \bibinfo {author} {\bibfnamefont {A.}~\bibnamefont {Stefanovska}},
  \ and\ \bibinfo {author} {\bibfnamefont {P.~V.~E.}\ \bibnamefont
  {McClintock}},\ }\bibfield  {title} {\enquote {\bibinfo {title} {Neuronal
  synchrony during anesthesia: {A} thalamocortical model},}\ }\href@noop {}
  {\bibfield  {journal} {\bibinfo  {journal} {Biophys. J.}\ }\textbf {\bibinfo
  {volume} {95}},\ \bibinfo {pages} {2722--2727} (\bibinfo {year}
  {2008})}\BibitemShut {NoStop}%
\bibitem [{\citenamefont {Bhowmik}\ and\ \citenamefont
  {Shanahan}(2012)}]{BhowmikShanahan12}%
  \BibitemOpen
  \bibfield  {author} {\bibinfo {author} {\bibfnamefont {D.}~\bibnamefont
  {Bhowmik}}\ and\ \bibinfo {author} {\bibfnamefont {M.}~\bibnamefont
  {Shanahan}},\ }\bibfield  {title} {\enquote {\bibinfo {title} {How well do
  oscillator models capture the behaviour of biological neurons?}}\ }in\
  \href@noop {} {\emph {\bibinfo {booktitle} {The 2012 International Joint
  Conference on Neural Networks (IJCNN)}}}\ (\bibinfo {year} {2012})\ pp.\
  \bibinfo {pages} {1--8}\BibitemShut {NoStop}%
\bibitem [{\citenamefont {Mirollo}\ and\ \citenamefont
  {Strogatz}(1990)}]{MirolloStrogatz90}%
  \BibitemOpen
  \bibfield  {author} {\bibinfo {author} {\bibfnamefont {R.}~\bibnamefont
  {Mirollo}}\ and\ \bibinfo {author} {\bibfnamefont {S.}~\bibnamefont
  {Strogatz}},\ }\bibfield  {title} {\enquote {\bibinfo {title}
  {Synchronization of pulse-coupled biological oscillators},}\ }\href {\doibase
  10.1137/0150098} {\bibfield  {journal} {\bibinfo  {journal} {SIAM J. Appl.
  Math.}\ }\textbf {\bibinfo {volume} {50}},\ \bibinfo {pages} {1645--1662}
  (\bibinfo {year} {1990})}\BibitemShut {NoStop}%
\bibitem [{\citenamefont {Filatrella}, \citenamefont {Nielsen},\ and\
  \citenamefont {Pedersen}(2008)}]{FilatrellaEtAl08}%
  \BibitemOpen
  \bibfield  {author} {\bibinfo {author} {\bibfnamefont {G.}~\bibnamefont
  {Filatrella}}, \bibinfo {author} {\bibfnamefont {A.~H.}\ \bibnamefont
  {Nielsen}}, \ and\ \bibinfo {author} {\bibfnamefont {N.~F.}\ \bibnamefont
  {Pedersen}},\ }\bibfield  {title} {\enquote {\bibinfo {title} {Analysis of a
  power grid using a {K}uramoto-like model},}\ }\href@noop {} {\bibfield
  {journal} {\bibinfo  {journal} {Eur. Phys. J. B}\ }\textbf {\bibinfo {volume}
  {61}},\ \bibinfo {pages} {485--491} (\bibinfo {year} {2008})}\BibitemShut
  {NoStop}%
\bibitem [{\citenamefont {Nishikawa}\ and\ \citenamefont
  {Motter}(2015{\natexlab{a}})}]{NishikawaMotter2015}%
  \BibitemOpen
  \bibfield  {author} {\bibinfo {author} {\bibfnamefont {T.}~\bibnamefont
  {Nishikawa}}\ and\ \bibinfo {author} {\bibfnamefont {A.~E.}\ \bibnamefont
  {Motter}},\ }\bibfield  {title} {\enquote {\bibinfo {title} {Comparative
  analysis of existing models for power-grid synchronization},}\ }\href
  {\doibase 10.1088/1367-2630/17/1/015012} {\bibfield  {journal} {\bibinfo
  {journal} {New J. Phys.}\ }\textbf {\bibinfo {volume} {17}} (\bibinfo {year}
  {2015}{\natexlab{a}}),\ 10.1088/1367-2630/17/1/015012}\BibitemShut {NoStop}%
\bibitem [{\citenamefont {Watanabe}\ and\ \citenamefont
  {Strogatz}(1994)}]{WatanabeStrogatz94}%
  \BibitemOpen
  \bibfield  {author} {\bibinfo {author} {\bibfnamefont {S.}~\bibnamefont
  {Watanabe}}\ and\ \bibinfo {author} {\bibfnamefont {S.~H.}\ \bibnamefont
  {Strogatz}},\ }\bibfield  {title} {\enquote {\bibinfo {title} {Constants of
  motion for superconducting {J}osephson arrays},}\ }\href@noop {} {\bibfield
  {journal} {\bibinfo  {journal} {Physica D}\ }\textbf {\bibinfo {volume}
  {74}},\ \bibinfo {pages} {197 -- 253} (\bibinfo {year} {1994})}\BibitemShut
  {NoStop}%
\bibitem [{\citenamefont {Wiesenfeld}, \citenamefont {Colet},\ and\
  \citenamefont {Strogatz}(1998)}]{WiesenfeldEtAl98}%
  \BibitemOpen
  \bibfield  {author} {\bibinfo {author} {\bibfnamefont {K.}~\bibnamefont
  {Wiesenfeld}}, \bibinfo {author} {\bibfnamefont {P.}~\bibnamefont {Colet}}, \
  and\ \bibinfo {author} {\bibfnamefont {S.~H.}\ \bibnamefont {Strogatz}},\
  }\bibfield  {title} {\enquote {\bibinfo {title} {Frequency locking in
  {J}osephson arrays: {C}onnection with the {K}uramoto model},}\ }\href
  {\doibase 10.1103/PhysRevE.57.1563} {\bibfield  {journal} {\bibinfo
  {journal} {Phys. Rev. E}\ }\textbf {\bibinfo {volume} {57}},\ \bibinfo
  {pages} {1563--1569} (\bibinfo {year} {1998})}\BibitemShut {NoStop}%
\bibitem [{\citenamefont {Kuramoto}(1984)}]{Kuramoto84}%
  \BibitemOpen
  \bibfield  {author} {\bibinfo {author} {\bibfnamefont {Y.}~\bibnamefont
  {Kuramoto}},\ }\href {\doibase 10.1007/978-3-642-69689-3} {\emph {\bibinfo
  {title} {Chemical {O}scillations, {W}aves, and {T}urbulence}}},\ \bibinfo
  {series} {Springer Series in Synergetics}, Vol.~\bibinfo {volume} {19}\
  (\bibinfo  {publisher} {Springer-Verlag},\ \bibinfo {address} {Berlin},\
  \bibinfo {year} {1984})\ pp.\ \bibinfo {pages} {viii+156}\BibitemShut
  {NoStop}%
\bibitem [{\citenamefont {Strogatz}(2000)}]{Strogatz00}%
  \BibitemOpen
  \bibfield  {author} {\bibinfo {author} {\bibfnamefont {S.~H.}\ \bibnamefont
  {Strogatz}},\ }\bibfield  {title} {\enquote {\bibinfo {title} {From
  {K}uramoto to {C}rawford: {E}xploring the onset of synchronization in
  populations of coupled oscillators},}\ }\href {\doibase
  10.1016/S0167-2789(00)00094-4} {\bibfield  {journal} {\bibinfo  {journal}
  {Physica D}\ }\textbf {\bibinfo {volume} {143}},\ \bibinfo {pages} {1--20}
  (\bibinfo {year} {2000})}\BibitemShut {NoStop}%
\bibitem [{\citenamefont {Pikovsky}, \citenamefont {Rosenblum},\ and\
  \citenamefont {Kurths}(2001)}]{PikovskyEtAl01}%
  \BibitemOpen
  \bibfield  {author} {\bibinfo {author} {\bibfnamefont {A.}~\bibnamefont
  {Pikovsky}}, \bibinfo {author} {\bibfnamefont {M.}~\bibnamefont {Rosenblum}},
  \ and\ \bibinfo {author} {\bibfnamefont {J.}~\bibnamefont {Kurths}},\
  }\href@noop {} {\emph {\bibinfo {title} {{Synchronization: {A} {U}niversal
  {C}oncept in {N}onlinear {S}ciences}}}}\ (\bibinfo  {publisher} {Cambridge
  University Press},\ \bibinfo {address} {Cambridge},\ \bibinfo {year}
  {2001})\BibitemShut {NoStop}%
\bibitem [{\citenamefont {Acebr\'on}\ \emph {et~al.}(2005)\citenamefont
  {Acebr\'on}, \citenamefont {Bonilla}, \citenamefont {P\'erez~Vicente},
  \citenamefont {Ritort},\ and\ \citenamefont {Spigler}}]{AcebronEtAl05}%
  \BibitemOpen
  \bibfield  {author} {\bibinfo {author} {\bibfnamefont {J.~A.}\ \bibnamefont
  {Acebr\'on}}, \bibinfo {author} {\bibfnamefont {L.~L.}\ \bibnamefont
  {Bonilla}}, \bibinfo {author} {\bibfnamefont {C.~J.}\ \bibnamefont
  {P\'erez~Vicente}}, \bibinfo {author} {\bibfnamefont {F.}~\bibnamefont
  {Ritort}}, \ and\ \bibinfo {author} {\bibfnamefont {R.}~\bibnamefont
  {Spigler}},\ }\bibfield  {title} {\enquote {\bibinfo {title} {The {K}uramoto
  model: {A} simple paradigm for synchronization phenomena},}\ }\href {\doibase
  10.1103/RevModPhys.77.137} {\bibfield  {journal} {\bibinfo  {journal} {Rev.
  Mod. Phys.}\ }\textbf {\bibinfo {volume} {77}},\ \bibinfo {pages} {137--185}
  (\bibinfo {year} {2005})}\BibitemShut {NoStop}%
\bibitem [{\citenamefont {Osipov}, \citenamefont {Kurths},\ and\ \citenamefont
  {Zhou}(2007)}]{OsipovEtAl07}%
  \BibitemOpen
  \bibfield  {author} {\bibinfo {author} {\bibfnamefont {G.~V.}\ \bibnamefont
  {Osipov}}, \bibinfo {author} {\bibfnamefont {J.}~\bibnamefont {Kurths}}, \
  and\ \bibinfo {author} {\bibfnamefont {C.}~\bibnamefont {Zhou}},\ }\href
  {\doibase 10.1007/978-3-540-71269-5} {\emph {\bibinfo {title}
  {Synchronization in {O}scillatory {N}etworks}}},\ Springer Series in
  Synergetics\ (\bibinfo  {publisher} {Springer},\ \bibinfo {address}
  {Berlin},\ \bibinfo {year} {2007})\ p.\ \bibinfo {pages} {37c}\BibitemShut
  {NoStop}%
\bibitem [{\citenamefont {Arenas}\ \emph {et~al.}(2008)\citenamefont {Arenas},
  \citenamefont {Diaz-Guilera}, \citenamefont {Kurths}, \citenamefont
  {Moreno},\ and\ \citenamefont {Zhou}}]{ArenasEtAl08}%
  \BibitemOpen
  \bibfield  {author} {\bibinfo {author} {\bibfnamefont {A.}~\bibnamefont
  {Arenas}}, \bibinfo {author} {\bibfnamefont {A.}~\bibnamefont
  {Diaz-Guilera}}, \bibinfo {author} {\bibfnamefont {J.}~\bibnamefont
  {Kurths}}, \bibinfo {author} {\bibfnamefont {Y.}~\bibnamefont {Moreno}}, \
  and\ \bibinfo {author} {\bibfnamefont {C.}~\bibnamefont {Zhou}},\ }\bibfield
  {title} {\enquote {\bibinfo {title} {Synchronization in complex networks},}\
  }\href {\doibase 10.1016/j.physrep.2008.09.002} {\bibfield  {journal}
  {\bibinfo  {journal} {Phys. Rep.}\ }\textbf {\bibinfo {volume} {469}},\
  \bibinfo {pages} {93--153} (\bibinfo {year} {2008})}\BibitemShut {NoStop}%
\bibitem [{\citenamefont {D{\"o}rfler}\ and\ \citenamefont
  {Bullo}(2014)}]{DorflerBullo14}%
  \BibitemOpen
  \bibfield  {author} {\bibinfo {author} {\bibfnamefont {F.}~\bibnamefont
  {D{\"o}rfler}}\ and\ \bibinfo {author} {\bibfnamefont {F.}~\bibnamefont
  {Bullo}},\ }\bibfield  {title} {\enquote {\bibinfo {title} {Synchronization
  in complex networks of phase oscillators: {A} survey},}\ }\href@noop {}
  {\bibfield  {journal} {\bibinfo  {journal} {Automatica}\ }\textbf {\bibinfo
  {volume} {50}},\ \bibinfo {pages} {1539 -- 1564} (\bibinfo {year}
  {2014})}\BibitemShut {NoStop}%
\bibitem [{\citenamefont {Rodrigues}\ \emph {et~al.}(2016)\citenamefont
  {Rodrigues}, \citenamefont {Peron}, \citenamefont {Ji},\ and\ \citenamefont
  {Kurths}}]{RodriguesEtAl16}%
  \BibitemOpen
  \bibfield  {author} {\bibinfo {author} {\bibfnamefont {F.~A.}\ \bibnamefont
  {Rodrigues}}, \bibinfo {author} {\bibfnamefont {T.~K.~D.}\ \bibnamefont
  {Peron}}, \bibinfo {author} {\bibfnamefont {P.}~\bibnamefont {Ji}}, \ and\
  \bibinfo {author} {\bibfnamefont {J.}~\bibnamefont {Kurths}},\ }\bibfield
  {title} {\enquote {\bibinfo {title} {The {K}uramoto model in complex
  networks},}\ }\href@noop {} {\bibfield  {journal} {\bibinfo  {journal} {Phys.
  Rep.}\ }\textbf {\bibinfo {volume} {610}},\ \bibinfo {pages} {1 -- 98}
  (\bibinfo {year} {2016})}\BibitemShut {NoStop}%
\bibitem [{\citenamefont {Ott}\ and\ \citenamefont
  {Antonsen}(2008)}]{OttAntonsen08}%
  \BibitemOpen
  \bibfield  {author} {\bibinfo {author} {\bibfnamefont {E.}~\bibnamefont
  {Ott}}\ and\ \bibinfo {author} {\bibfnamefont {T.~M.}\ \bibnamefont
  {Antonsen}},\ }\bibfield  {title} {\enquote {\bibinfo {title} {Low
  dimensional behavior of large systems of globally coupled oscillators},}\
  }\href {\doibase 10.1063/1.2930766} {\bibfield  {journal} {\bibinfo
  {journal} {Chaos}\ }\textbf {\bibinfo {volume} {18}},\ \bibinfo {pages}
  {037113, 6} (\bibinfo {year} {2008})}\BibitemShut {NoStop}%
\bibitem [{\citenamefont {Gottwald}(2015)}]{Gottwald15}%
  \BibitemOpen
  \bibfield  {author} {\bibinfo {author} {\bibfnamefont {G.~A.}\ \bibnamefont
  {Gottwald}},\ }\bibfield  {title} {\enquote {\bibinfo {title} {Model
  reduction for networks of coupled oscillators},}\ }\href@noop {} {\bibfield
  {journal} {\bibinfo  {journal} {Chaos}\ }\textbf {\bibinfo {volume} {25}},\
  \bibinfo {pages} {053111, 12} (\bibinfo {year} {2015})}\BibitemShut {NoStop}%
\bibitem [{\citenamefont {Gottwald}(2017)}]{Gottwald17}%
  \BibitemOpen
  \bibfield  {author} {\bibinfo {author} {\bibfnamefont {G.~A.}\ \bibnamefont
  {Gottwald}},\ }\bibfield  {title} {\enquote {\bibinfo {title} {Finite-size
  effects in a stochastic {K}uramoto model},}\ }\href {\doibase
  10.1063/1.5004618} {\bibfield  {journal} {\bibinfo  {journal} {Chaos}\
  }\textbf {\bibinfo {volume} {27}},\ \bibinfo {pages} {101103} (\bibinfo
  {year} {2017})}\BibitemShut {NoStop}%
\bibitem [{\citenamefont {Hancock}\ and\ \citenamefont
  {Gottwald}(2018)}]{HancockGottwald18}%
  \BibitemOpen
  \bibfield  {author} {\bibinfo {author} {\bibfnamefont {E.~J.}\ \bibnamefont
  {Hancock}}\ and\ \bibinfo {author} {\bibfnamefont {G.~A.}\ \bibnamefont
  {Gottwald}},\ }\bibfield  {title} {\enquote {\bibinfo {title} {Model
  reduction for {K}uramoto models with complex topologies},}\ }\href {\doibase
  10.1103/PhysRevE.98.012307} {\bibfield  {journal} {\bibinfo  {journal} {Phys.
  Rev. E}\ }\textbf {\bibinfo {volume} {98}},\ \bibinfo {pages} {012307}
  (\bibinfo {year} {2018})}\BibitemShut {NoStop}%
\bibitem [{\citenamefont {Smith}\ and\ \citenamefont
  {Gottwald}(2019)}]{SmithGottwald19}%
  \BibitemOpen
  \bibfield  {author} {\bibinfo {author} {\bibfnamefont {L.~D.}\ \bibnamefont
  {Smith}}\ and\ \bibinfo {author} {\bibfnamefont {G.~A.}\ \bibnamefont
  {Gottwald}},\ }\bibfield  {title} {\enquote {\bibinfo {title} {Chaos in
  networks of coupled oscillators with multimodal natural frequency
  distributions},}\ }\href {\doibase 10.1063/1.5109130} {\bibfield  {journal}
  {\bibinfo  {journal} {Chaos}\ }\textbf {\bibinfo {volume} {29}},\ \bibinfo
  {pages} {093127} (\bibinfo {year} {2019})}\BibitemShut {NoStop}%
\bibitem [{\citenamefont {Smith}\ and\ \citenamefont
  {Gottwald}(2020)}]{SmithGottwald20}%
  \BibitemOpen
  \bibfield  {author} {\bibinfo {author} {\bibfnamefont {L.~D.}\ \bibnamefont
  {Smith}}\ and\ \bibinfo {author} {\bibfnamefont {G.~A.}\ \bibnamefont
  {Gottwald}},\ }\bibfield  {title} {\enquote {\bibinfo {title} {Model
  reduction for the collective dynamics of globally coupled oscillators: From
  finite networks to the thermodynamic limit},}\ }\href {\doibase
  10.1063/5.0009790} {\bibfield  {journal} {\bibinfo  {journal} {Chaos}\
  }\textbf {\bibinfo {volume} {30}},\ \bibinfo {pages} {093107} (\bibinfo
  {year} {2020})}\BibitemShut {NoStop}%
\bibitem [{\citenamefont {Yue}, \citenamefont {Smith},\ and\ \citenamefont
  {Gottwald}(2020)}]{YueEtAl20}%
  \BibitemOpen
  \bibfield  {author} {\bibinfo {author} {\bibfnamefont {W.}~\bibnamefont
  {Yue}}, \bibinfo {author} {\bibfnamefont {L.~D.}\ \bibnamefont {Smith}}, \
  and\ \bibinfo {author} {\bibfnamefont {G.~A.}\ \bibnamefont {Gottwald}},\
  }\bibfield  {title} {\enquote {\bibinfo {title} {Model reduction for the
  {K}uramoto-{S}akaguchi model: The importance of nonentrained rogue
  oscillators},}\ }\href {\doibase 10.1103/PhysRevE.101.062213} {\bibfield
  {journal} {\bibinfo  {journal} {Phys. Rev. E}\ }\textbf {\bibinfo {volume}
  {101}},\ \bibinfo {pages} {062213} (\bibinfo {year} {2020})}\BibitemShut
  {NoStop}%
\bibitem [{\citenamefont {Menck}\ \emph {et~al.}(2014)\citenamefont {Menck},
  \citenamefont {Heitzig}, \citenamefont {Kurths},\ and\ \citenamefont
  {Joachim~Schellnhuber}}]{MenckEtAl14}%
  \BibitemOpen
  \bibfield  {author} {\bibinfo {author} {\bibfnamefont {P.~J.}\ \bibnamefont
  {Menck}}, \bibinfo {author} {\bibfnamefont {J.}~\bibnamefont {Heitzig}},
  \bibinfo {author} {\bibfnamefont {J.}~\bibnamefont {Kurths}}, \ and\ \bibinfo
  {author} {\bibfnamefont {H.}~\bibnamefont {Joachim~Schellnhuber}},\
  }\bibfield  {title} {\enquote {\bibinfo {title} {How dead ends undermine
  power grid stability},}\ }\href {\doibase 10.1038/ncomms4969} {\bibfield
  {journal} {\bibinfo  {journal} {Nat. Comm.}\ }\textbf {\bibinfo {volume}
  {5}},\ \bibinfo {pages} {3969} (\bibinfo {year} {2014})}\BibitemShut
  {NoStop}%
\bibitem [{\citenamefont {Erd{\H{o}}s}\ and\ \citenamefont
  {R{\'e}nyi}(1960)}]{ErdosRenyi60}%
  \BibitemOpen
  \bibfield  {author} {\bibinfo {author} {\bibfnamefont {P.}~\bibnamefont
  {Erd{\H{o}}s}}\ and\ \bibinfo {author} {\bibfnamefont {A.}~\bibnamefont
  {R{\'e}nyi}},\ }\bibfield  {title} {\enquote {\bibinfo {title} {On the
  evolution of random graphs},}\ }\href@noop {} {\bibfield  {journal} {\bibinfo
   {journal} {Publ. Math. Inst. Hung. Acad. Sci}\ }\textbf {\bibinfo {volume}
  {5}},\ \bibinfo {pages} {17--60} (\bibinfo {year} {1960})}\BibitemShut
  {NoStop}%
\bibitem [{\citenamefont {Barab{\'a}si}\ and\ \citenamefont
  {Albert}(1999)}]{BarabasiAlbert99}%
  \BibitemOpen
  \bibfield  {author} {\bibinfo {author} {\bibfnamefont {A.-L.}\ \bibnamefont
  {Barab{\'a}si}}\ and\ \bibinfo {author} {\bibfnamefont {R.}~\bibnamefont
  {Albert}},\ }\bibfield  {title} {\enquote {\bibinfo {title} {Emergence of
  scaling in random networks},}\ }\href {\doibase 10.1126/science.286.5439.509}
  {\bibfield  {journal} {\bibinfo  {journal} {Science}\ }\textbf {\bibinfo
  {volume} {286}},\ \bibinfo {pages} {509--512} (\bibinfo {year}
  {1999})}\BibitemShut {NoStop}%
\bibitem [{Note1()}]{Note1}%
  \BibitemOpen
  \bibinfo {note} {For each value of $K$, stationary states are found by
  solving $\protect \mathaccentV {dot}05F\phi _i = 0$ for $i=1,\protect \dots
  ,N$ in the full Kuramoto model (\ref {eq:full_KM}) using the multidimensional
  Newton root finding method. If a stationary state is found, then its
  stability is checked by determining the eigenvalues of the Jacobian. We
  perform bisection in $K$ until $K_g$ is found within a tolerance of
  $10^{-4}$.}\BibitemShut {Stop}%
\bibitem [{\citenamefont {Peron}\ \emph {et~al.}(2019)\citenamefont {Peron},
  \citenamefont {Messias F.~de Resende}, \citenamefont {Mata}, \citenamefont
  {Rodrigues},\ and\ \citenamefont {Moreno}}]{PeronEtAl19}%
  \BibitemOpen
  \bibfield  {author} {\bibinfo {author} {\bibfnamefont {T.}~\bibnamefont
  {Peron}}, \bibinfo {author} {\bibfnamefont {B.}~\bibnamefont {Messias F.~de
  Resende}}, \bibinfo {author} {\bibfnamefont {A.~S.}\ \bibnamefont {Mata}},
  \bibinfo {author} {\bibfnamefont {F.~A.}\ \bibnamefont {Rodrigues}}, \ and\
  \bibinfo {author} {\bibfnamefont {Y.}~\bibnamefont {Moreno}},\ }\bibfield
  {title} {\enquote {\bibinfo {title} {Onset of synchronization of {K}uramoto
  oscillators in scale-free networks},}\ }\href
  {https://link.aps.org/doi/10.1103/PhysRevE.100.042302} {\bibfield  {journal}
  {\bibinfo  {journal} {Phys. Rev. E}\ }\textbf {\bibinfo {volume} {100}},\
  \bibinfo {pages} {042302} (\bibinfo {year} {2019})}\BibitemShut {NoStop}%
\bibitem [{\citenamefont {Sakaguchi}\ and\ \citenamefont
  {Kuramoto}(1986)}]{SakaguchiKuramoto86}%
  \BibitemOpen
  \bibfield  {author} {\bibinfo {author} {\bibfnamefont {H.}~\bibnamefont
  {Sakaguchi}}\ and\ \bibinfo {author} {\bibfnamefont {Y.}~\bibnamefont
  {Kuramoto}},\ }\bibfield  {title} {\enquote {\bibinfo {title} {{A Soluble
  Active Rotater Model Showing Phase Transitions via Mutual Entertainment}},}\
  }\href {\doibase 10.1143/PTP.76.576} {\bibfield  {journal} {\bibinfo
  {journal} {Prog. Theor. Phys.}\ }\textbf {\bibinfo {volume} {76}},\ \bibinfo
  {pages} {576--581} (\bibinfo {year} {1986})}\BibitemShut {NoStop}%
\bibitem [{\citenamefont {Yang}, \citenamefont {Nishikawa},\ and\ \citenamefont
  {Motter}(2017{\natexlab{a}})}]{YangEtAl17}%
  \BibitemOpen
  \bibfield  {author} {\bibinfo {author} {\bibfnamefont {Y.}~\bibnamefont
  {Yang}}, \bibinfo {author} {\bibfnamefont {T.}~\bibnamefont {Nishikawa}}, \
  and\ \bibinfo {author} {\bibfnamefont {A.~E.}\ \bibnamefont {Motter}},\
  }\bibfield  {title} {\enquote {\bibinfo {title} {Small vulnerable sets
  determine large network cascades in power grids},}\ }\href
  {https://science.sciencemag.org/content/358/6365/eaan3184} {\bibfield
  {journal} {\bibinfo  {journal} {Science}\ }\textbf {\bibinfo {volume} {358}}
  (\bibinfo {year} {2017}{\natexlab{a}})}\BibitemShut {NoStop}%
\bibitem [{\citenamefont {Nishikawa}\ and\ \citenamefont
  {Motter}(2015{\natexlab{b}})}]{NishikawaMotter15}%
  \BibitemOpen
  \bibfield  {author} {\bibinfo {author} {\bibfnamefont {T.}~\bibnamefont
  {Nishikawa}}\ and\ \bibinfo {author} {\bibfnamefont {A.~E.}\ \bibnamefont
  {Motter}},\ }\bibfield  {title} {\enquote {\bibinfo {title} {Comparative
  analysis of existing models for power-grid synchronization},}\ }\href
  {\doibase 10.1088/1367-2630/17/1/015012} {\bibfield  {journal} {\bibinfo
  {journal} {New J. Phys.}\ }\textbf {\bibinfo {volume} {17}},\ \bibinfo
  {pages} {015012} (\bibinfo {year} {2015}{\natexlab{b}})}\BibitemShut
  {NoStop}%
\bibitem [{\citenamefont {Yang}, \citenamefont {Nishikawa},\ and\ \citenamefont
  {Motter}(2017{\natexlab{b}})}]{YangEtAl17b}%
  \BibitemOpen
  \bibfield  {author} {\bibinfo {author} {\bibfnamefont {Y.}~\bibnamefont
  {Yang}}, \bibinfo {author} {\bibfnamefont {T.}~\bibnamefont {Nishikawa}}, \
  and\ \bibinfo {author} {\bibfnamefont {A.~E.}\ \bibnamefont {Motter}},\
  }\bibfield  {title} {\enquote {\bibinfo {title} {Vulnerability and
  cosusceptibility determine the size of network cascades},}\ }\href {\doibase
  10.1103/PhysRevLett.118.048301} {\bibfield  {journal} {\bibinfo  {journal}
  {Phys. Rev. Lett.}\ }\textbf {\bibinfo {volume} {118}},\ \bibinfo {pages}
  {048301} (\bibinfo {year} {2017}{\natexlab{b}})}\BibitemShut {NoStop}%
\bibitem [{\citenamefont {Olmi}\ \emph {et~al.}(2014)\citenamefont {Olmi},
  \citenamefont {Navas}, \citenamefont {Boccaletti},\ and\ \citenamefont
  {Torcini}}]{OlmiEtAl14}%
  \BibitemOpen
  \bibfield  {author} {\bibinfo {author} {\bibfnamefont {S.}~\bibnamefont
  {Olmi}}, \bibinfo {author} {\bibfnamefont {A.}~\bibnamefont {Navas}},
  \bibinfo {author} {\bibfnamefont {S.}~\bibnamefont {Boccaletti}}, \ and\
  \bibinfo {author} {\bibfnamefont {A.}~\bibnamefont {Torcini}},\ }\bibfield
  {title} {\enquote {\bibinfo {title} {Hysteretic transitions in the kuramoto
  model with inertia},}\ }\href {\doibase 10.1103/PhysRevE.90.042905}
  {\bibfield  {journal} {\bibinfo  {journal} {Phys. Rev. E}\ }\textbf {\bibinfo
  {volume} {90}},\ \bibinfo {pages} {042905} (\bibinfo {year}
  {2014})}\BibitemShut {NoStop}%
\end{thebibliography}%

\end{document}